%% file: main.tex
\documentclass[a4paper,fleqn]{cas-dc}
\usepackage[authoryear,longnamesfirst]{natbib}
\usepackage{placeins}
\usepackage{longtable,booktabs,algorithm,tabularx,multirow,algpseudocode,graphicx,xspace}
\providecommand{\Description}[1]{}
\newcommand{\name}{CineAssist\xspace}

\begin{document}
\renewcommand{\printorcid}{}
\let\WriteBookmarks\relax
\def\floatpagepagefraction{1}
\def\textpagefraction{.001}
\shorttitle{CineAssist}
\shortauthors{Y. He et~al.}
\title[mode=title]{Designing with Cinematic Language: An Intent-Driven Approach to Human-AI Video Clip Co-Creation}

\author[1]{Yi He}
\ead{heyi_11@tongji.edu.cn}
\affiliation[1]{organization={iDVX Lab, College of Design and Innovation, Tongji University},
            city={Shanghai},
            country={China}}

\author[1,2]{Zhengxi Yu}
\ead{2452633@tongji.edu.cn}

\affiliation[2]{organization={Innovation Center for Intelligent Design, Shanghai Innovation Institute},
            city={Shanghai},
            country={China}}
\affiliation[3]{organization={Shanghai Research Institute for Intelligent Autonomous Systems},
            city={Shanghai},
            country={China}}
\author[1,2]{Feilin Han}
\ead{hanfeilin@tongji.edu.cn}

\author[1,3]{Chuer Chen}
\ead{chuerchen@tongji.edu.cn}

\author[1,2,3]{Shengqi Dang}
\ead{dangsq123@163.com}

\author[1,2]{Nan Cao}
\cormark[1]
\ead{nan.cao@gmail.com}
\cortext[1]{Nan Cao is the corresponding author}

\begin{abstract}
Generative video models have made video creation more accessible, yet creators still struggle to translate high-level narrative intentions into coherent cinematic expressions. Existing systems primarily rely on prompt refinement or expose isolated camera controls, leaving a gap between \textit{what to express} and \textit{how to film it}. We present \name, a human-AI co-creation system that treats cinematic language as an interactive design medium connecting creative intentions with executable video controls. We construct a Cinematic Language Design Space and an annotated dataset from 107 filmmaking sources and 1,436 film clips, linking communicative intentions to cinematic parameter values. CineAssist combines a multi-agent workflow for intent alignment and strategy recommendation with an interactive cinematic editor and a context-sensitive preference learner for user-controlled personalization. We evaluate the system through three studies examining end-to-end usability, cinematic quality, and preference evolution. Results show that CineAssist facilitates efficient, controllable, and personalized intent-driven cinematic video creation.
\end{abstract}
\maketitle

\input{Sections/01-Introduction}
\input{Sections/02-RelatedWork}
\input{Sections/03-FormativeStudy}

\input{Sections/04-System}

\input{Sections/05-DesignSpace}

\input{Sections/06-Agents}
\input{Sections/07-Evaluation}
\input{Sections/08-Discussion}

\input{Sections/09-Conclusion}

\section*{CRediT authorship contribution statement}
\textbf{Yi He:} Writing -- original draft, Writing -- review \& editing, Visualization, Validation, Software, Methodology, Investigation, Conceptualization.
\textbf{Zhengxi Yu:} Writing -- original draft, Software, Methodology, Investigation.
\textbf{Feilin Han:} Supervision, Writing -- review \& editing, Conceptualization.
\textbf{Chuer Chen:} Data curation, Conceptualization.
\textbf{Shengqi Dang:} Data curation, Conceptualization.
\textbf{Nan Cao:} Supervision, Writing -- review \& editing, Conceptualization.

\section*{Declaration of Generative AI and AI-assisted technologies in the writing process}
During the preparation of this work, the authors used ChatGPT to polish the language. After using this tool, the authors reviewed and edited the content as needed and take full responsibility for the content of the published article.

\section*{Declaration of competing interest}
The authors declare that they have no known competing financial interests or personal relationships that could have appeared to influence the work reported in this paper.

\section*{Acknowledgements}
Nan Cao is the corresponding author. This work was supported by the Innovation Ecosystem Development Program (International Science and Technology Cooperation Project) of the Science and Technology Commission of Shanghai Municipality (Grant No. 26510713800).

\clearpage
\bibliographystyle{cas-model2-names}
\bibliography{main}
\end{document}

%% file: Sections/01-Introduction.tex
\section{Introduction}
Cinematic language is a fundamental medium for visual storytelling. Through choices of camera position, viewpoint, focal length, movement, composition, lighting, and rhythm, creators can transform the same narrative content into different emotional and perceptual experiences~\citep{bordwell2019film,brown2021cinematography,block2021visualstory}. Yet designing such expressions remains largely dependent on professional knowledge: creators must translate high-level goals such as ``make the character feel isolated'' or ``create tension'' into coordinated choices across multiple cinematic dimensions, and iteratively refine these choices through repeated preview and adjustment~\citep{brown2021cinematography,collagevis2024}. This expertise barrier makes cinematic storytelling hard for non-experts to access and motivates creativity-support tools that help users explore and refine creative alternatives~\citep{shneiderman2007creativity}.

Generative AI has substantially lowered the barrier to producing visual content. Text-to-video and image-guided video models allow users to generate dynamic scenes from natural-language descriptions and visual references~\citep{singer2023makeavideo,ho2022imagenvideo,blattmann2023align}. However, these systems primarily address \textit{what} to generate rather than \textit{how} to express it through cinematic language. Recent work has introduced controls for camera trajectories, motion, viewpoints, and previsualization workflows~\citep{wang2023videocomposer,wang2024motionctrl,directavideo2024,he2025cameractrl,liu2024chatcam,chen2024cinepregen,wei2025cinevision,hu2026previzwhiz}, but these approaches typically expose individual controls or organize production assets without providing a unified representation that connects communicative intents to coordinated cinematic choices. Consequently, users still face a semantic gap between expressing \textit{what they want to communicate} and specifying \textit{how it should be filmed}, often relying on prompt iteration and stochastic generation to discover satisfactory results~\citep{liu2022opal,wang2023reprompt}.

This gap presents three challenges for human-AI cinematic creation: (1) cinematic expression is inherently \emph{multi-dimensional}: parameters such as composition, viewpoint, optics, movement, lighting, and rhythm interact rather than functioning as independent controls; (2) users naturally communicate in terms of \emph{high-level intents}, whereas generative models require concrete and coordinated parameter choices. Translating between these representations therefore requires more than prompt generation. It requires an intermediate representation that is both semantically meaningful to users and actionable for generation; (3) cinematic design is \emph{iterative and personal}: users need to compare alternative realizations, refine individual choices, and develop recurring preferences, while existing generative workflows provide limited support for maintaining such knowledge across tasks.
\begin{figure*}[t]

  \centering
  \includegraphics[width=0.96\textwidth]{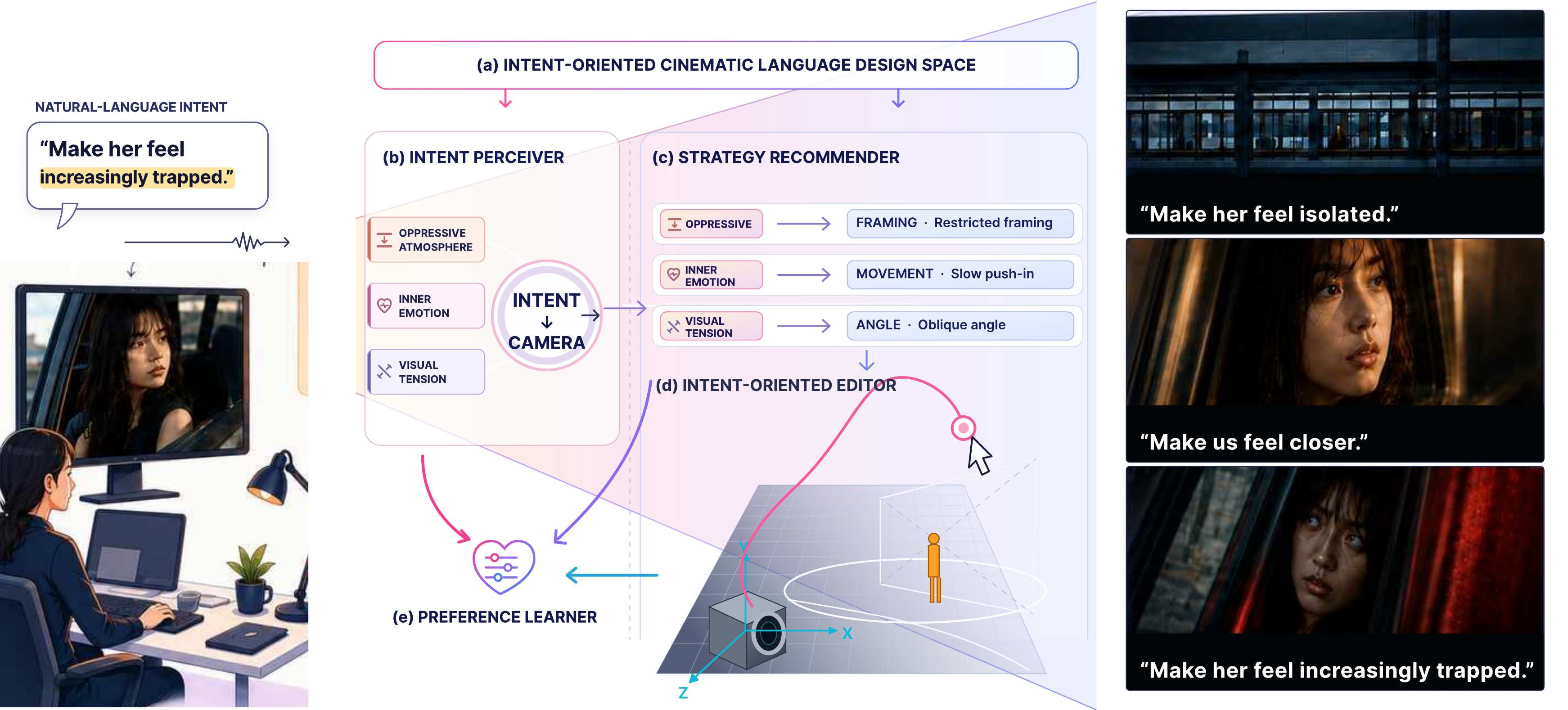}
  \caption{Overview of \name{}. Guided by (a) the cinematic language design space, (b) the Intent Perceiver clarifies vague requests into explicit intentions, and (c) the Strategy Recommender translates them into executable Shot Scripts. (d) The editor instantiates and refines the resulting camera configurations in 3D, while (e) the Preference Learner learns reusable preferences from creators' choices and edits. Together, these modules produce controllable cinematic rewrites that communicate the intended effect.}
  \Description{\name{} workflow from creative intent to generated video. A user provides a reference image and the request ``Make her feel increasingly trapped.'' The Intent Perceiver aligns the request with communicative intentions in the design space, including oppressive atmosphere, inner emotion, and visual tension. The Strategy Recommender maps these intentions to framing, camera movement, and angle choices. The user refines the resulting shot in a three-dimensional editor, while the Preference Learner captures choices and edits for future tasks. The final column shows videos generated from the same reference image under different creative intentions.}
  \label{fig:teaser}

\end{figure*}
We present \name, a human-AI collaborative system for cinematic language-driven video creation. \name treats cinematic language as an intermediate design space between users' communicative intents and generative video control. As shown in Figure~\ref{fig:teaser}, the system first establishes a structured \emph{Cinematic Language Design Space} that connects communicative intents with cinematic parameter values. Building on this representation, a multi-agent system supports the creation workflow: the \emph{Intent Perceiver} translates underspecified requests into user-confirmed Intent Briefs, while the \emph{Strategy Recommender} generates and validates alternative cinematic realizations. An interactive \emph{Cinematic Editor} then allows users to inspect and refine these designs directly, with the resulting parameter changes translated into generation controls. Finally, the \emph{Preference Learner} accumulates evidence from users' interaction histories and retrieves contextually relevant preferences for confirmation in subsequent tasks, enabling personalization without automatically overriding the user's current intent. Our contributions are:

\begin{itemize}

\item \textbf{A multi-agent workflow for translating creative intent into cinematic expression.}
We present \name, a multi-agent workflow that transforms users' high-level creative intents into alternative, validated cinematic strategies and connects them to interactive editing and generative video creation. By making the translation from ``what to express'' to ``how to film'' explicit and editable, the system supports users in inspecting, comparing, and iteratively refining cinematic expression rather than relying on one-shot prompt-based generation.

\item \textbf{A structured Cinematic Language Design Space and annotated dataset.}
We develop a structured design space that organizes communicative intents and their associated cinematic parameter values, providing a shared semantic representation for both users and agents. To construct and ground this space, we collect and annotate a large-scale cinematic dataset that captures the relationships between communicative intents and cinematic parameter choices. The resulting design space and dataset provide an empirical foundation for retrieving, reasoning about, and exploring alternative cinematic realizations, bridging high-level expressive goals and actionable cinematic controls without prescribing deterministic mappings.

\item \textbf{A context-sensitive preference learning mechanism for user-controlled personalization.}
We introduce a preference learner that accumulates evidence from users' cinematic decisions across comparable tasks, distinguishes recurring preferences from isolated or conflicting choices, and retrieves relevant preferences for user confirmation in subsequent creation tasks. This enables reusable personalization while preserving user control over when learned preferences are applied.

\end{itemize}

%% file: Sections/02-RelatedWork.tex
\section{Related Work}
We situate our work at the intersection of three areas: generative video creation and control, computational cinematography, and human--AI co-creation for visual media. Together, these areas motivate our focus on connecting cinematic intent, cinematographic design, and interactive generation.

\subsection{Generative AI for Video Creation}

Diffusion-based video generation has advanced rapidly in visual quality, duration, and temporal coherence. Make-A-Video extends pretrained image models with spatial--temporal modules~\citep{singer2023makeavideo}, while Imagen Video produces high-definition clips through cascaded spatial and temporal super-resolution~\citep{ho2022imagenvideo}. Video LDM adapts latent diffusion to video through temporal layers and aligned upsamplers~\citep{blattmann2023align}, and Stable Video Diffusion highlights the importance of large-scale data curation and video pretraining~\citep{blattmann2023svd}. More recent architectures model longer and more diverse sequences: CogVideoX employs spatial--temporal latent compression~\citep{yang2024cogvideox}, Sora represents videos as space--time patches~\citep{openai2024sora}, and Lumiere generates an entire clip through a space--time U-Net to improve motion coherence~\citep{bartal2024lumiere}. Together, these models provide increasingly capable foundations for single-shot video generation.

Recent methods further expose explicit control over generated motion. AnimateDiff introduces transferable motion modules~\citep{guo2024animatediff}, while VideoComposer incorporates conditions such as sketches, depth maps, and motion vectors~\citep{wang2023videocomposer}. DragNUWA controls object motion through user-specified trajectories~\citep{yin2023dragnuwa}, and Boximator uses bounding boxes to define object locations and movement~\citep{wang2024boximator}. CameraCtrl conditions generation on camera poses~\citep{he2025cameractrl}, whereas MotionCtrl separately controls camera and object motion~\citep{wang2024motionctrl}. These methods improve controllability but generally require users to directly specify poses, trajectories, regions, or motion parameters, offering limited guidance on which controls best communicate an intended narrative or emotional effect.

HCI research has integrated AI into broader video-authoring workflows. QuickCut uses transcripts to support narrated-video composition~\citep{truong2016quickcut}. Generative Disco supports music-responsive visual generation~\citep{liu2024generativedisco}, while ReelFramer transforms news articles into editable scripts and storyboards~\citep{wang2024reelframer}. LAVE combines an LLM-based agent with direct video-editing controls~\citep{wang2024lave}. These systems primarily support content organization, asset generation, or editing operations. In contrast, \name{} addresses the intermediate translation from an underspecified communicative intent to cinematic expression by proposing alternative cinematic strategies and mapping a selected strategy to interpretable, editable shot parameters.

\subsection{Computational Cinematography}
Computational cinematography has developed extensive methods for representing and automating camera control. Early systems encoded cinematographic idioms and reusable camera behaviors for virtual filming~\citep{drucker1995camdroid,he1996virtual,christianson1996declarative}, followed by constraint-based methods for computing camera configurations~\citep{bares2000constraintcam} and compact representations such as Toric Space for interactive viewpoint and composition control~\citep{lino2015toric}. Subsequent work has addressed shot continuity and editing~\citep{galvane2015continuity,leake2017computational}, learned camera behaviors from film examples~\citep{jiang2020example,jiang2021keyframing}, and incorporated physical constraints for robotic cinematography~\citep{bonatti2020autonomous}. Together, this body of work provides a strong computational foundation for camera placement, framing, motion, and continuity~\citep{christie2008cameracontrol,ronfard2021film}.

However, these approaches typically begin with an explicit scene representation, camera objective, reference, or low-level constraint. They therefore offer limited support for the earlier creative step of translating an ambiguous narrative intent into alternative cinematic strategies, or for helping users understand how parameter choices affect communicative meaning. Our work builds on computational cinematography by introducing an intent-centered layer that connects communicative goals to alternative cinematic strategies and makes their parameter-level realization inspectable and revisable.

\subsection{Human-AI Co-Creation for Visual Content}

Human--AI creativity tools increasingly support generation as an iterative process of proposing, inspecting, and revising alternatives rather than one-shot creation. Prompt-based systems assist users in exploring and refining textual specifications through generated suggestions, retrieved examples, and visual feedback~\citep{brade2023promptify,feng2024promptmagician,wang2024promptcharm}. Other systems complement language with direct or spatial interaction, allowing users to manipulate visual concepts, spatial prompts, or generative strokes~\citep{chung2023promptpaint,marquardt2025imaginationvellum}. For sequential media, interactive environments further expose editable intermediate representations, such as visual stories, animation code, and semantic video segments~\citep{antony2024id8,tseng2024keyframer,leake2024chunkyedit}. Together, these systems demonstrate the value of making intermediate representations explicit and editable, enabling users to inspect and refine generated content rather than treating model outputs as fixed results.

However, existing approaches largely address prompt formulation, visual manipulation, or content-level editing, leaving the translation from high-level creative intent to cinematic expression comparatively underexplored. Prompt-oriented tools help users specify or refine \textit{what} to generate, while direct-manipulation systems often assume that users already know \textit{how} a desired effect should be realized. Few systems explicitly connect intent clarification with cinematography-grounded strategy generation and interactive parameter-level editing, while also adapting to users' recurring choices without obscuring or overriding their control. Our work addresses this gap by making the transition from ``what to express'' to ``how to film'' explicit, editable, and iterative within a unified human--AI workflow.

%% file: Sections/03-FormativeStudy.tex
\section{Formative Study}
\label{sec:formative-study}
We conducted a formative study to examine how video creators use generative AI in video production, characterize their workflows, and identify key challenges.

\subsection{Participants}
We recruited 10 participants: 6 experienced directors or filmmaking practitioners (E1--E6; 4--8 years of experience, M = 5.5) and 4 design students without formal directing training (P1--P4). All participants had at least one year's experience of using AI-based video generation tools. We recruited participants through referrals from faculty at a film academy and direct outreach to local studios. Each participant received \$7 compensation. Table~\ref{tab:participants} summarizes participants' filmmaking experience, prior study participation, and commonly used AI tools.

\begin{table*}[t]
\centering
\caption{Participant demographics and commonly used GenAI tools.}
\Description{Demographic information for the 10 formative-study participants. Six experienced participants include directors, a writer-director, and artificial-intelligence filmmaking practitioners with four to eight years of filmmaking experience. Four participants are design students with one to three years of experience. The table also lists age, gender, background, and commonly used generative artificial intelligence tools, including Runway, Kling, Vidu, Pika, Dreamina, Stable Diffusion, and ComfyUI.}
\label{tab:participants}
\small
\setlength{\tabcolsep}{5pt}
\renewcommand{\arraystretch}{1.15}

\begin{tabularx}{\textwidth}{
    @{}
    l
    l
    c
    c
    l
    c
    >{\raggedright\arraybackslash}X
    @{}
}
\toprule
ID & Group & Age & Gender & Background & YoE & GenAI tools used \\
\midrule

E1 & Experienced & 29 & M & Director
   & 5 years & Runway, Kling, Jimeng AI (Dreamina) \\

E2 & Experienced & 34 & M & Writer-director
   & 4 years & Vidu, Runway, Kling, Jimeng AI (Dreamina) \\

E3 & Experienced & 31 & F & Director
   & 5 years & Jimeng AI (Dreamina), ComfyUI, Kling \\

E4 & Experienced & 28 & M & AI filmmaking practitioner
   & 5 years & Vidu, Kling \\

E5 & Experienced & 36 & M & Director
   & 6 years & Jimeng AI (Dreamina), Runway, Pika, Kling \\

E6 & Experienced & 39 & M & AI filmmaking practitioner
   & 8 years & Stable Diffusion, ComfyUI, Runway, Kling \\

\midrule

P1 & Student & 22 & F & Design student
   & 2 years & Vidu, Pika, Kling \\

P2 & Student & 23 & F & Design student
   & 2 years & Jimeng AI (Dreamina), Runway \\

P3 & Student & 24 & F & Design student
   & 1 year & Vidu, Kling \\

P4 & Student & 22 & M & Design student
   & 3 years & Jimeng AI (Dreamina), Stable Diffusion \\

\bottomrule
\end{tabularx}
\end{table*}

\subsection{Study Procedure}
We conducted semi-structured interviews lasting about 60 minutes via video calls. The interview protocol focused on three areas: (1) participants' workflows and experiences with AI video-generation tools; (2) their decision-making practices when designing and refining  video shots; and (3) challenges and support needs throughout the design, generation, and revision process. The semi-structured format ensured consistency across interviews while allowing participants to elaborate on their individual experiences. We transcribed the interviews and conducted open coding to identify recurring themes, grouping related observations across participants.

\begin{figure*}[t]
    \centering
\includegraphics[width=1\linewidth]{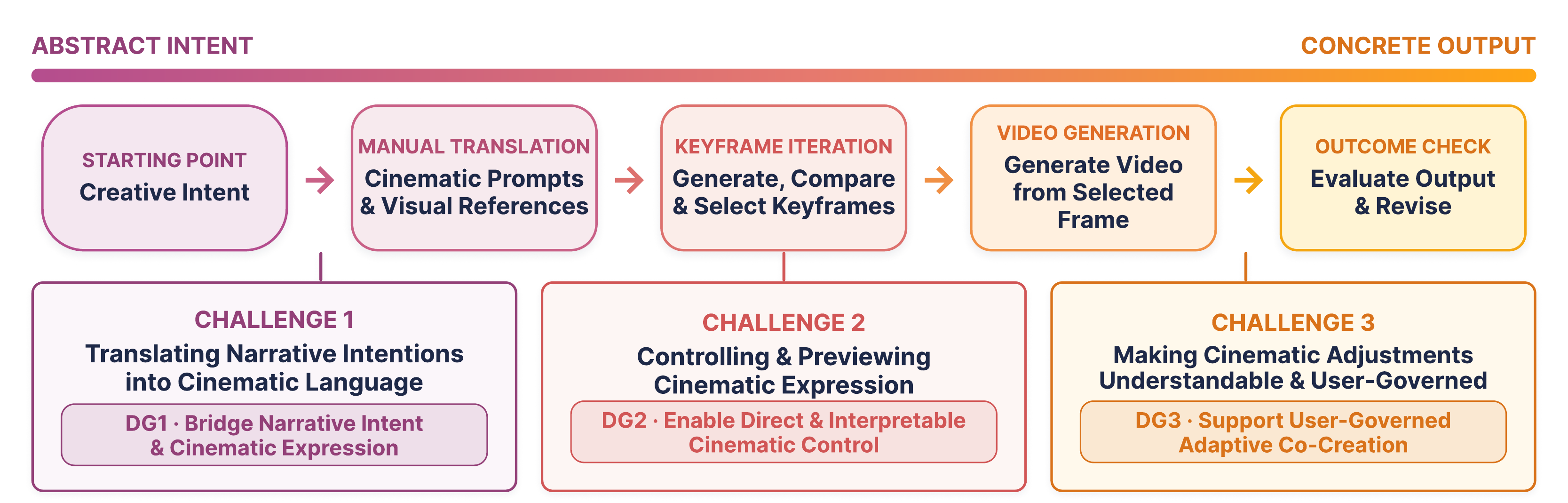}
    \caption{Current GenAI video-production workflow, the three challenges identified in the formative study, and their corresponding design goals.}
    \Description{Five-stage generative artificial intelligence video workflow and associated design challenges. The workflow proceeds from a creative intent, through manual translation into prompts and visual references, iterative keyframe generation and selection, video generation, and evaluation of the output. Three challenges appear beneath the workflow: translating narrative intentions into cinematic language, controlling and previewing cinematic expression, and making adjustments semantically understandable and user-governed. These lead respectively to three design goals concerning intent translation, interpretable control, and adaptive co-creation.}
    \label{fig:Findings}
\end{figure*}

\subsection{Findings}
As illustrated in Figure~\ref{fig:Findings}, participants described an iterative GenAI video-production workflow: (1) articulating a narrative or emotional intent, (2) translating it into prompts and references, (3) generating and comparing keyframes, (4) selecting a candidate for video generation, and (5) evaluating and refining the result. When the generated framing, movement, or visual effect diverged from the intended expression, participants revised the prompts or references and repeated the cycle. Across both participant groups, three challenges emerged.

\paragraph{\textbf{Challenge 1: Translating Intents into Cinematic Language}}
Participants could often articulate \textit{what} a scene should communicate but struggled to specify \textit{how} it should be filmed. Experienced practitioners noted that GenAI could follow explicit camera instructions but often failed to capture the narrative function of those choices within a scene~(E1--E2, E4). As one practitioner explained, \textit{``AI knows how to push in when instructed, but it does not understand what a push-in means in this particular scene''}~(E1).

This gap was particularly evident for high-level intents such as conveying oppression or loneliness. Broad prompts often produced exaggerated or generic treatments, while achieving a more specific cinematic expression required detailed prompts, carefully selected references, or substantial filmmaking expertise~(E2, E4). The challenge was even greater for design students without formal cinematography training: they could describe the intended emotion or meaning but often lacked the vocabulary to map it to camera, composition, or other cinematic choices~(P1--P3). Thus, participants faced a fundamental translation problem between \textit{what to express} and \textit{how to film it}.

\paragraph{\textbf{Challenge 2: Controlling and Previewing Cinematic Expression}}
Even when participants had a clear cinematic idea, realizing and evaluating it through prompt-based generation remained difficult. Participants repeatedly generated and compared alternatives to achieve the desired result~(E1--E4, P1--P3); one practitioner reported, \textit{``I can spend an entire afternoon generating variations just to get one scene right''}~(E3).

Prompts could often specify coarse attributes such as shot scale, but provided less reliable control over composition, camera angle, character--camera relationships, and movement~(E1). This made it difficult to preserve spatial relationships or predict how a specified cinematic design would appear in the generated video. Some experienced practitioners had experimented with 3D director's-desk tools for arranging cameras and subjects, yet still relied on generation to validate the resulting design. Consequently, cinematic decisions remained tightly coupled to trial-and-error generation rather than being directly inspectable and controllable.

\paragraph{\textbf{Challenge 3: Understanding and Governing Cinematic Adjustments}}
Participants wanted more direct control over cinematic parameters, but emphasized that exposing professional controls alone was insufficient. Less-experienced creators often knew the desired expressive effect but did not know which parameter to change or what expressive consequence that change would produce. As one student explained, \textit{``I may know that I want the character to feel isolated, but I do not know whether I should change the framing, composition, or camera movement''}~(P4).

Participants therefore wanted cinematic controls to be connected to their expressive implications, so that they could understand why alternative parameter choices might strengthen or weaken an intended effect. They preferred this guidance to remain contextual rather than prescriptive: the system should explain possible cinematic consequences while leaving the final decision to the user.
Participants also wanted the system to recognize recurring creative choices across interactions~(E2, E5, P1--P3). For example, one practitioner suggested that repeated choices in dialogue scenes should gradually become part of the film's style~(E5). At the same time, participants emphasized that learned preferences should remain visible and user-governed, with control over when and where they are applied and the ability to inspect, modify, or reject them.

\subsection{Design Goals}

The findings reveal three gaps in current GenAI video-creation workflows: the gap between narrative intent and cinematic expression (C1), the limited direct control and preview of cinematic decisions (C2), and the lack of semantically grounded yet user-governed adaptation (C3). We therefore derived three design goals for CineAssist.

\begin{itemize}
    \item \textbf{DG1: Bridge Narrative Intent and Cinematic Expression.} The system should help creators articulate and refine what they want to express, and translate these intents into interpretable cinematic strategies. Rather than prescribing a single solution, it should expose alternative cinematic expressions and allow creators to select the one that best fits their intent.
    \item \textbf{DG2: Enable Direct and Interpretable Cinematic Control.} The system should allow creators to directly manipulate key cinematic elements and preview their consequences before generation. Such interaction should make framing, viewpoint, spatial relationships, and camera movement observable and adjustable, reducing reliance on prompt-based trial and error.
    \item \textbf{DG3: Support User-Governed Adaptive Co-Creation.} The system should connect cinematic adjustments to their semantic and expressive consequences while learning from users' selections, edits, and feedback. Adaptation should remain transparent and user-controlled, with learned preferences visible, editable, reversible, and bounded by user-defined scope.
\end{itemize}

Together, these goals frame CineAssist as a Human-AI co-creation system that supports creators across three interconnected activities: translating intent into cinematic expression, directly shaping cinematic decisions, and progressively adapting the collaboration to individual creative preferences.

%% file: Sections/04-System.tex

\section{System Design}

Building on the design goals derived from our formative study, we developed \name, an intent-driven multi-agent system that bridges creators' high-level intentions and executable cinematic designs. We first present the overall system architecture, user interface, and interaction workflow, followed by the technical details of its key components in subsequent sections.

\begin{figure*}[t]
    \centering
    \includegraphics[width=1\linewidth]{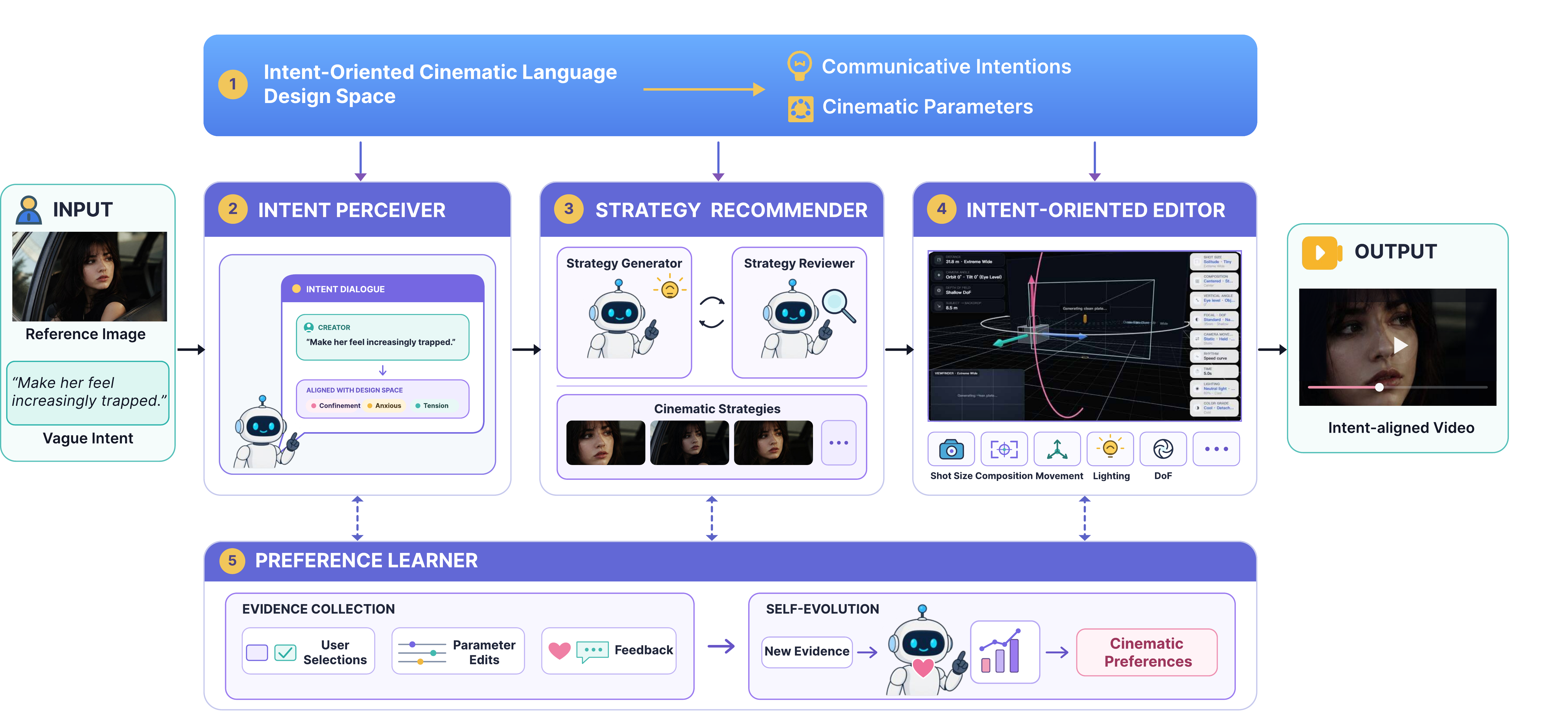}
\caption{Overview of \name. The system consists of five interconnected modules: the Intent-Oriented Cinematic Language Design Space, Intent Perceiver, Strategy Recommender, Intent-Oriented Editor, and Preference Learner.}
\Description{System architecture of \name. A reference image and vague creative intent enter the Intent Perceiver, which aligns the request with communicative intentions in the cinematic language design space. The Strategy Generator and Reviewer create and validate alternative cinematic strategies. A selected strategy enters the Intent-Oriented Editor, where users adjust shot size, composition, movement, lighting, depth of field, and rhythm to produce an intent-aligned video. User selections, parameter edits, and feedback also flow to the Preference Learner, which develops reusable preferences for subsequent tasks.}
    \label{fig:system}
\end{figure*}
\subsection{System Overview}
Rather than requiring creators to specify low-level camera parameters, \name is designed to use cinematic language as an intermediate design space to translate intentions into alternative strategies, enable direct and interpretable refinement, and progressively adapt to creators' preferences. As shown in Figure~\ref{fig:system}, \name comprises five interconnected modules that support intent-driven cinematic creation: \textit{\textbf{(1) Intent-Oriented Cinematic Language Design Space}}, which organizes creative intentions and their corresponding cinematic language and parameters; \textit{\textbf{(2) Intent Perceiver}}, an agent that interprets creators' descriptions and aligns them with structured intentions in the Design Space~\textbf{(DG1)}; \textit{\textbf{(3) Strategy Recommender}}, a multi-agent system comprising a Strategy Generator and a Strategy Reviewer that translates aligned intentions into alternative cinematic strategies and parameter configurations~\textbf{(DG1)}; \textit{\textbf{(4) Intent-Oriented Editor}}, an interactive tool that enables creators to directly manipulate and refine cinematic parameters while understanding their expressive implications~\textbf{(DG2, DG3)}; and \textit{\textbf{(5) Preference Learner}}, a multi-agent system that learns creators' recurring cinematic preferences from their interactions and feedback~\textbf{(DG3)}.

\begin{figure*}[t]
    \centering
\includegraphics[width=1\linewidth]{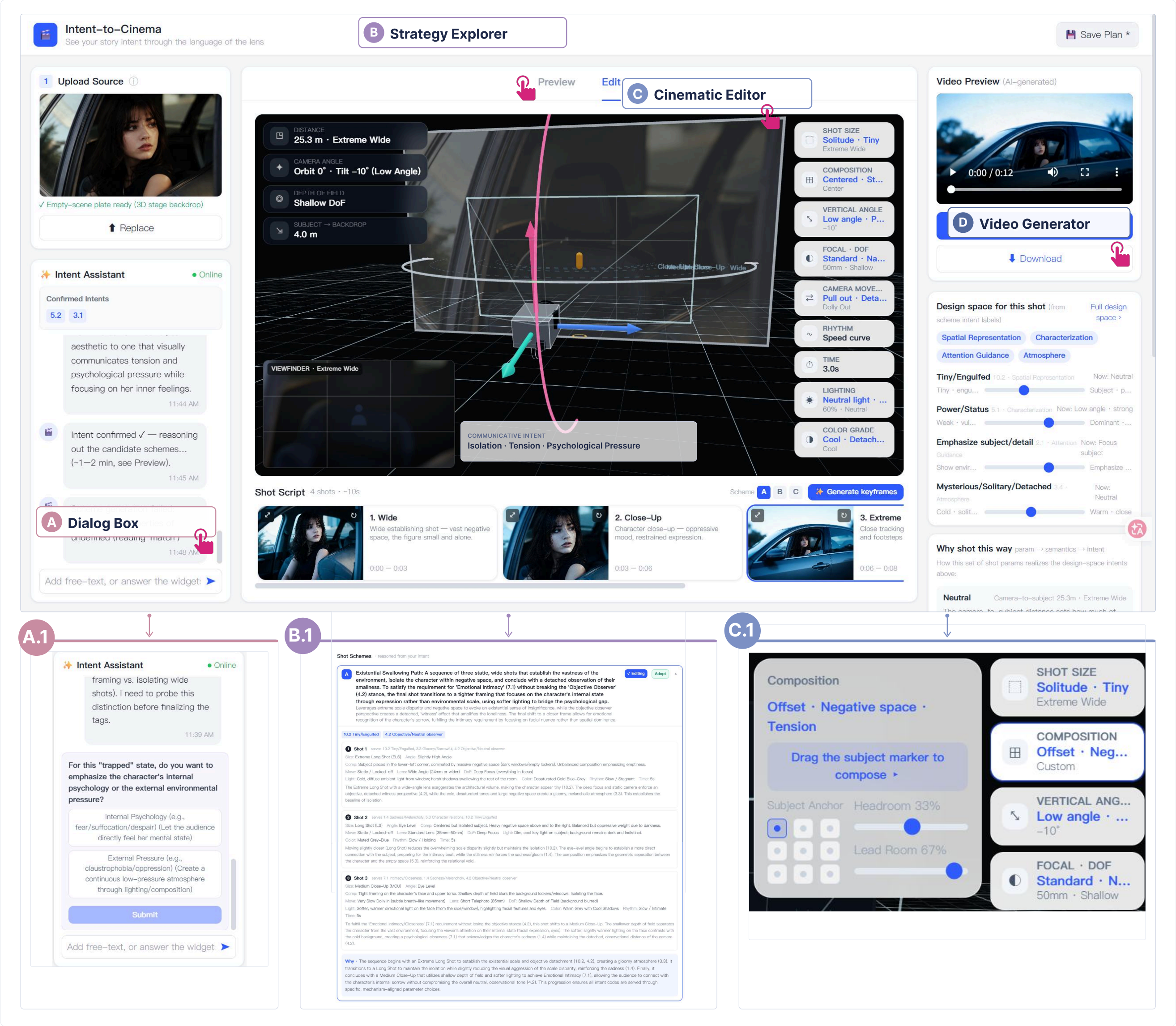}
    \caption{The CineAssist interface comprises (A) the Dialog Box for clarifying and confirming intentions, (B) the Strategy Explorer for comparing alternative strategies and Shot Scripts, (C) the Cinematic Editor for 3D refinement, and (D) the Video Generator for generating and previewing results. Insets A.1, B.1, and C.1 show representative interactions in greater detail.}
    \Description{The \name interface. The Dialog Box on the left supports reference-image upload and intent clarification. The Strategy Explorer at the top center presents alternative multi-shot strategies. The Cinematic Editor occupies the main central area, combining a three-dimensional scene, camera trajectory, frame preview, and cinematic parameter controls. The Video Generator on the right displays generated results. Enlarged insets illustrate intent dialogue, strategy selection, and direct manipulation of camera framing.}
    \label{fig:ui}
\end{figure*}

\subsection{User Interface}
To make the system modules accessible during creation, \name organizes the user interface into four interconnected views (Figure~\ref{fig:ui}): (1) the \emph{Dialog Box} (Fig.~\ref{fig:ui}(a)) for articulating and confirming creative intentions, (2) the \emph{Strategy Explorer} (Figure~\ref{fig:ui}(b)) for exploring alternative cinematic strategies, (3) the \emph{Cinematic Editor} (Figure~\ref{fig:ui}(c)) for refining cinematic designs, and (4) the \emph{Video Generator} (Figure~\ref{fig:ui}(d)) for generating and evaluating video results. The views share a common cinematic design representation, maintaining the correspondence among intent, strategy, shot configuration, and generated video. This shared representation allows creators to move between design stages and revisit earlier decisions without restarting the workflow.

\paragraph{\textbf{Dialog Box.}}
The Dialog Box provides the primary interface for intent alignment. Creators upload a reference image and describe their desired narrative or emotional effect in natural language. The system interprets the description against the Cinematic Language Design Space and identifies ambiguities that may affect subsequent cinematic decisions. Targeted questions and interactive choices help creators clarify their intent, after which the resulting \emph{Intent Brief} is presented for confirmation. This confirmed representation becomes the input to strategy exploration.

\paragraph{\textbf{Strategy Explorer.}}
The Strategy Explorer presents alternative cinematic strategies derived from the confirmed intent. Each strategy summarizes how the same creative goal can be realized through different configurations of shot size, composition, viewpoint, camera movement, optics, lighting, and other cinematic properties. Strategies can include multiple shots, enabling creators to compare alternative cinematic treatments at the shot-sequence level and select one for further refinement.

\paragraph{\textbf{Cinematic Editor.}}
The Cinematic Editor provides a synchronized 3D workspace for inspecting and refining the selected cinematic strategy. Creators can directly manipulate camera--subject relationships, composition, viewpoint, camera movement, lighting, and other cinematic parameters through spatial controls and parameter panels. Changes are synchronized across the 3D scene, parameter values, and shot preview, allowing creators to immediately inspect their effects.
The editor further supports intent-driven refinement by connecting parameter adjustments to their expressive meanings. For each parameter, the interface indicates the adjustment direction that is expected to better support the confirmed intent, as shown in Figure~\ref{fig:ui}(C.1). As creators modify the configuration, the editor also displays the communicative intentions jointly conveyed by the current parameter settings, helping them assess whether the evolving design remains consistent with the intended cinematic effect.

Cinematic parameters can be edited either spatially in the 3D workspace or through the side panel. The spatial controls are centered on the selected subject, represented by a yellow marker, allowing creators to configure the camera relative to the subject. Creators can \textbf{adjust shot size} by dragging the green arrow toward or away from the subject; \textbf{adjust camera angle} by dragging the pink control vertically or the blue control horizontally to orbit around the subject; and \textbf{adjust composition} by selecting one of nine regions in the rule-of-thirds grid, which automatically repositions the camera. For \textbf{camera movement}, creators can record successive camera positions and orientations as keyframes to construct an editable trajectory.

Other cinematic properties are edited through the side panel. Creators can adjust \textbf{focal length} to control field of view and perspective compression, \textbf{depth of field} to control focus distance and the range of sharpness, \textbf{lighting} to configure light position, intensity, hardness, and color, and \textbf{color} to modify tone, saturation, and color temperature. Creators can also adjust \textbf{rhythm} through a speed curve that controls the acceleration and deceleration of camera movement, and specify \textbf{shot duration}, which is synchronized with the Shot Script to maintain the pacing of the overall sequence. All changes are reflected immediately in the corresponding scene and preview.

\subsection{Interaction Workflow}

\name{} supports an iterative workflow that moves from \textbf{\emph{intent alignment}} to \textbf{\emph{strategy exploration}}, \textbf{\emph{cinematic refinement}}, and \textbf{\emph{video generation}}. The creator first provides a reference image and describes the desired narrative or emotional effect in the \textbf{\emph{Dialog Box}}. The \textbf{\emph{Intent Perceiver}} uses the Design Space to interpret the request, asks targeted questions when necessary, and produces a user-confirmed \textbf{\emph{Intent Brief}}. The \textbf{\emph{Strategy Recommender}} then generates and validates multiple cinematic strategies that realize the confirmed intent through different parameter configurations, which the creator compares in the \textbf{\emph{Strategy Explorer}}. After selecting a strategy, the creator refines its shot-level composition, viewpoint, camera movement, and other cinematic parameters in the \textbf{\emph{Cinematic Editor}}, with semantic guidance linking parameter changes to the intended effect. The finalized design is passed to the \textbf{\emph{Video Generator}} to produce the video, after which the creator can return to any relevant design decision and regenerate as needed. Across sessions, the \textbf{\emph{Preference Learner}} accumulates repeated, context-consistent interaction evidence and retrieves relevant preferences for subsequent tasks, while keeping their application visible and subject to user confirmation. The following sections describe the Design Space and the key agent components that support this workflow.

%% file: Sections/05-DesignSpace.tex
\section{Intent-Oriented Cinematic Language Design Space}

\begin{figure*}[t]
    \centering
    \includegraphics[width=\linewidth]{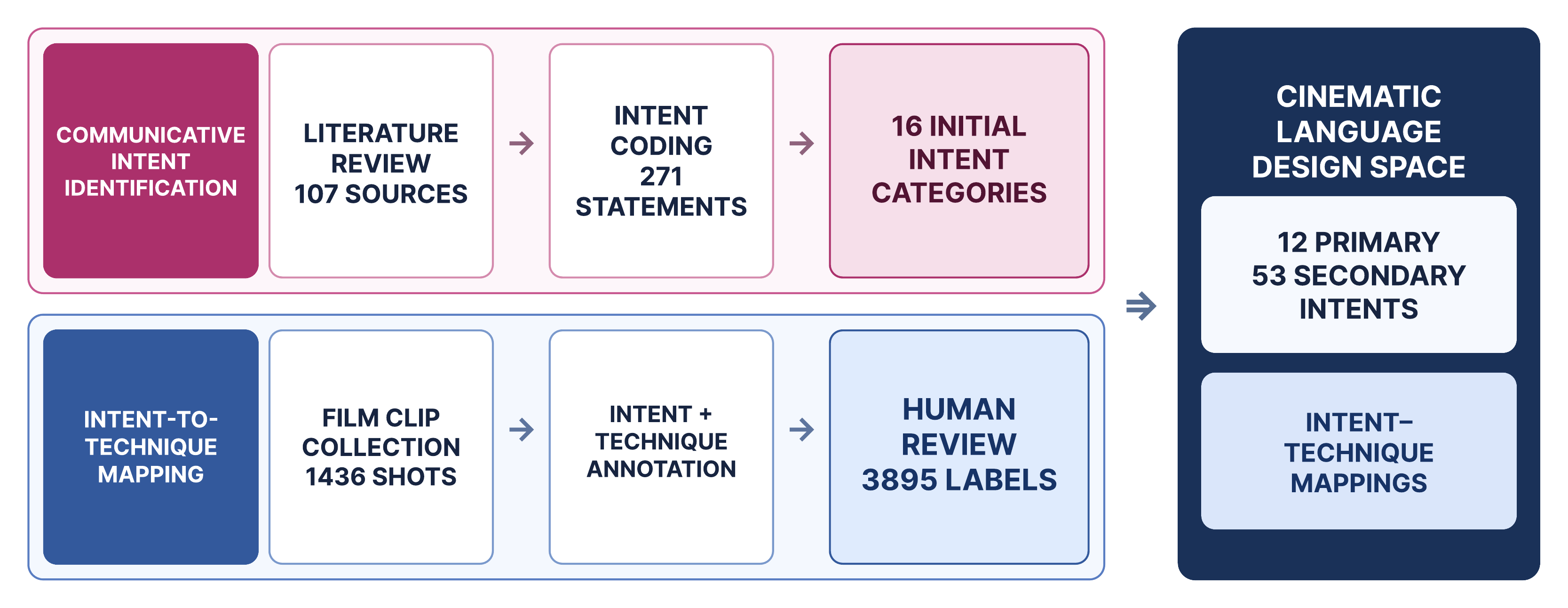}
    \caption{Construction of the cinematic language design space. A literature review identifies what creators use cinematic language to communicate, while film-clip annotation characterizes how these intents are realized through cinematic techniques.}
    \Description{Two-part process for constructing the cinematic language design space. The literature-analysis branch collects 107 sources, extracts intention and technique statements, and organizes them into initial categories. The film-analysis branch collects 1,436 film shots, filters them for intent and technique clarity, and produces 3,895 intention-technique annotations. The two branches are integrated to create mappings between communicative intentions and cinematic parameters.}
    \label{fig:method}
\end{figure*}

As revealed by our formative study, creators often struggle to translate high-level, underspecified intents into concrete cinematic decisions. To bridge this gap, we developed the Intent-Oriented Cinematic Language Design Space, which connects creators' expressive intents with concrete cinematic techniques and parameters, serving as a shared knowledge foundation for intent perception, strategy reasoning, and intent-aware editing in \name. The Design Space consists of two complementary components: (1) communicative intents that can be expressed through cinematic language, and (2) a set of commonly used cinematic parameters for realizing these intents. As shown in Figure~\ref{fig:method}, we construct these two components through a literature review and film-clip annotation, respectively, as detailed in the following sections.

\begin{figure}[width=\columnwidth,pos=ht]
    \centering
\includegraphics[width=1.1\linewidth]{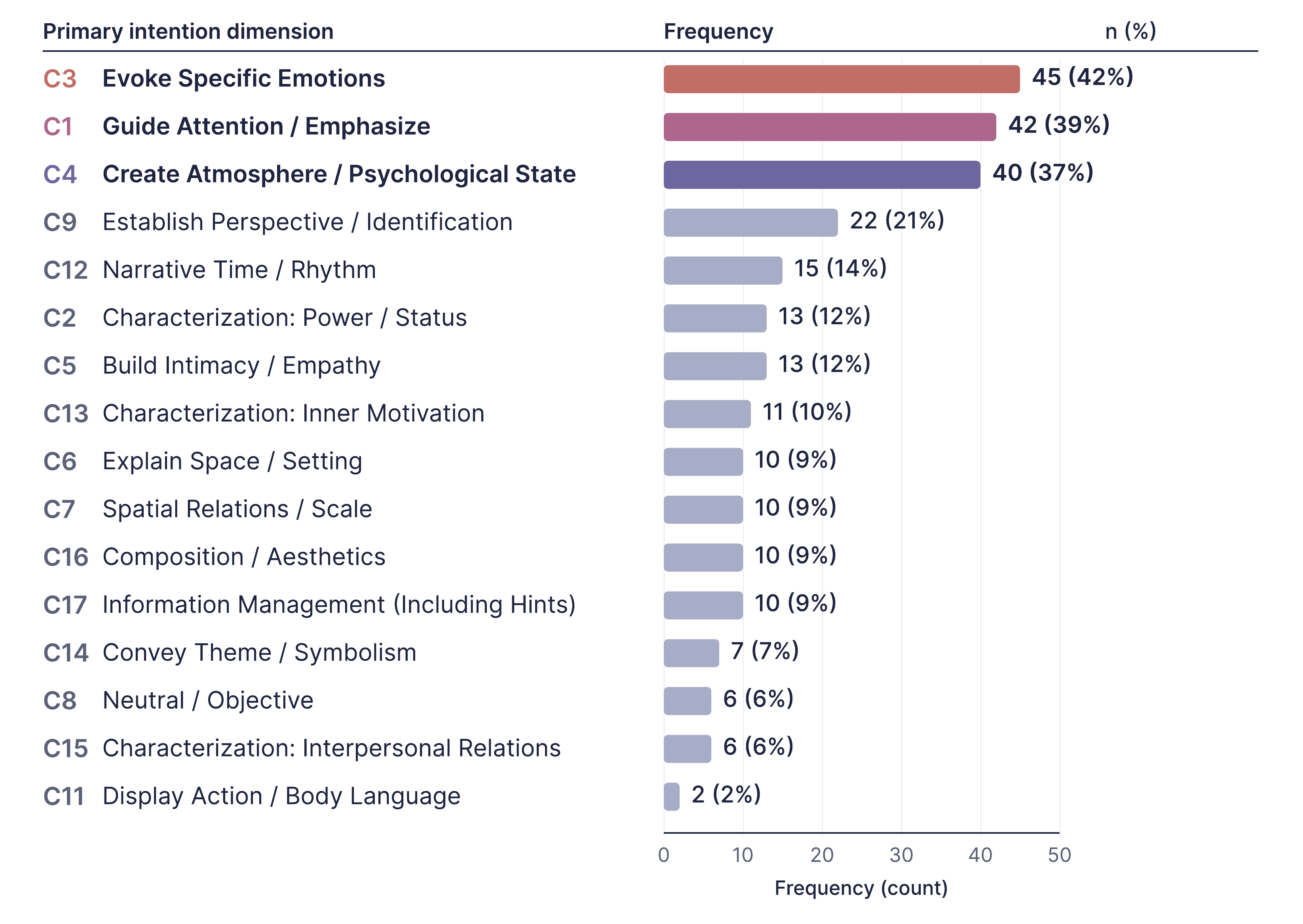}
    \caption{Frequency of the 16 initial intent categories identified from the literature.}
    \Description{Horizontal bar chart of 16 intention categories identified in the literature. Evoking specific emotions is the most frequent category with 45 sources, followed by guiding attention or emphasis with 42 and creating atmosphere or psychological state with 40. Establishing perspective or identification occurs 22 times, while most remaining categories occur between 2 and 15 times. Displaying action or body language is the least frequent category with 2 occurrences.}
    \label{fig:intents}
\end{figure}

\subsection{Identifying Communicative Intents}
We first investigated what cinematic language can communicate and the vocabulary through which these intents are expressed. We reviewed 107 authoritative sources across five traditions: practitioner accounts~\citep{brown2021cinematography,mascelli1998fivecs,block2021visualstory}, film studies~\citep{bordwell2019film,bordwell1985narration,salt1974statistical}, cognitive and affective research~\citep{hochberg1978perception,tan1996emotion,cutting2011quicker}, visual semiotics and rhetoric~\citep{barthes1977rhetoric,kress2006readingimages,messaris1997visualpersuasion}, and film-education curricula~\citep{bowen2013grammaroftheshot,mercado2010filmmakerseye}. We mined the collected text corpus to extract statements linking cinematic techniques to communicative effects and statistically analyzed their frequency and semantic co-occurrence. This analysis yielded 16 initial high-level communicative intent categories (Figure~\ref{fig:intents}), covering purposes such as evoking emotions, directing attention, and creating atmosphere. It also identified nine cinematic parameter categories that constitute the vocabulary of cinematic language and provide the controllable means for expressing these intents: shot scale, composition, camera angle and viewpoint, camera movement, focal length, depth of field, lighting, color, and rhythm (Table~\ref{tab:cinematic-techniques}). Together, these intent categories and cinematic dimensions form the initial foundation of the Design Space, which we subsequently validated and refined through film-clip annotation.

\begin{table*}[ht]
\centering
\caption{Cinematic parameter categories and representative values used in the annotation codebook.}
\Description{Annotation codebook containing nine cinematic parameter dimensions and representative values. The dimensions are shot scale, composition, camera angle and viewpoint, camera movement, focal length, depth of field, lighting, color, and rhythm. Representative values range from extreme-wide to extreme-close framing; centered, rule-of-thirds, symmetric, and imbalanced composition; high, eye-level, and low angles; static and moving camera techniques; different lens and focus choices; lighting and color treatments; and stagnant, gentle, steady, or urgent rhythms.}
\label{tab:cinematic-techniques}
\small
\renewcommand{\arraystretch}{1.12}
\begin{tabularx}{\textwidth}{
    >{\raggedright\arraybackslash}p{0.22\textwidth}
    >{\raggedright\arraybackslash}X
}
\toprule
\textbf{Dimension} & \textbf{Representative Values} \\
\midrule
Shot scale & Extreme wide; wide; full; medium; medium close-up; close-up; extreme close-up \\
Composition & Centered; symmetry; rule of thirds; leading lines; negative space; foreground occlusion \\
Camera angle and viewpoint & Eye-level; low angle; high angle; overhead; subjective/POV; objective observation \\
Camera movement & Static; push-in; pull-out; pan; tilt; tracking; orbit; handheld \\
Focal length & Wide-angle; standard; telephoto \\
Depth of field & Deep; medium; shallow \\
Lighting & Natural; soft; hard; low-key; high-key; chiaroscuro; backlighting \\
Color & Warm; cool; low saturation; high saturation \\
Rhythm & Stagnant; gentle; steady; urgent  \\
\bottomrule
\end{tabularx}
\end{table*}

\subsection{Mapping Intents to Cinematic Techniques}

The literature review identified what cinematic language can communicate, but not how these intents are realized in practice. We therefore constructed and annotated a film-clip corpus to examine how communicative intents are expressed through observable cinematic techniques. The resulting annotations refined the initial intent taxonomy and provided the empirical basis for the intent--cinematic technique mappings in the Design Space.

\paragraph{\textbf{Clip Corpus.}}
We constructed a corpus of 1436 clips from 1005 films, sampled from Academy Award for Best Cinematography nominees and winners~\citep{oscars_database}, the first 200 titles in the IMDb Top 250~\citep{imdb_top250}, and excerpts from the official Movieclips YouTube channel~\citep{movieclips}, accessed in \textit{June 2026}. After removing duplicates, we selected the shortest one-to-three-shot sequence that formed a coherent narrative or affective moment and provided sufficient visual quality and local context for annotation. We excluded trailers, promotional re-edits, unrelated montages, and excerpts requiring substantial contextual information. Each clip served as the annotation unit and could receive multiple intent and cinematic-technique labels. The corpus is released at \{anonymous link\}.

\paragraph{\textbf{Clip Annotation.}}
We annotated each clip with its communicative intent(s) and corresponding cinematic parameter values to characterize how cinematic language realizes different intents. We employed an AI-assisted annotation pipeline. First, we used GPT-5.6~\citep{openai2026gpt56} to generate candidate intent and parameter labels for each clip, which were independently reviewed by another two multimodal large models, Qwen 3.7-plus~\citep{qwen2026qwen37plus} and Doubao-seed 2.1 Pro~\citep{doubao2026seed21pro}. Consistent labels were retained, while disagreements, uncertain values, and newly proposed intents were flagged for human review. Three human annotators then reviewed the flagged cases, resolving disagreements and refining intent labels when necessary. This process yielded 3895 instances in form of <intent, cinematic-parameter-values>. Among them, 2864 were consistently identified by all three models, 915 were identified by a subset of the models and subsequently reconciled through human review, and 116 added or determined through human review.

The annotations also allowed us to refine the literature-derived taxonomy against cinematic practice. We merged overlapping categories, separated distinct communicative goals, and added recurring intents absent from the initial framework. The resulting taxonomy comprises 12 primary intent families and 53 secondary intents, which, together with the annotated cinematic parameters, form the empirical foundation of the Design Space.

\paragraph{\textbf{Intent--Cinematic Mapping.}}
To characterize how communicative intents are realized through cinematic language, we analyzed the distribution and co-occurrence of cinematic parameter values across the nine parameter dimensions for each of the 53 secondary intents. We first identified frequently used parameter values within each intent and then examined their recurring combinations across dimensions. This analysis revealed multiple cinematic realizations for the same intent, as well as the contextual reuse of individual parameter values across different intents. Rather than assigning each intent a fixed parameter configuration, we retained these many-to-many associations to form the intent--cinematic mappings in the Design Space. These mappings provide the empirical knowledge basis for intent perception, strategy recommendation, and intent-aware editing in \name.

\subsection{The Resulting Design Space}
\label{sec:resulting-design-space}

\begin{figure*}[t]
    \centering
    \includegraphics[width=0.89\linewidth]{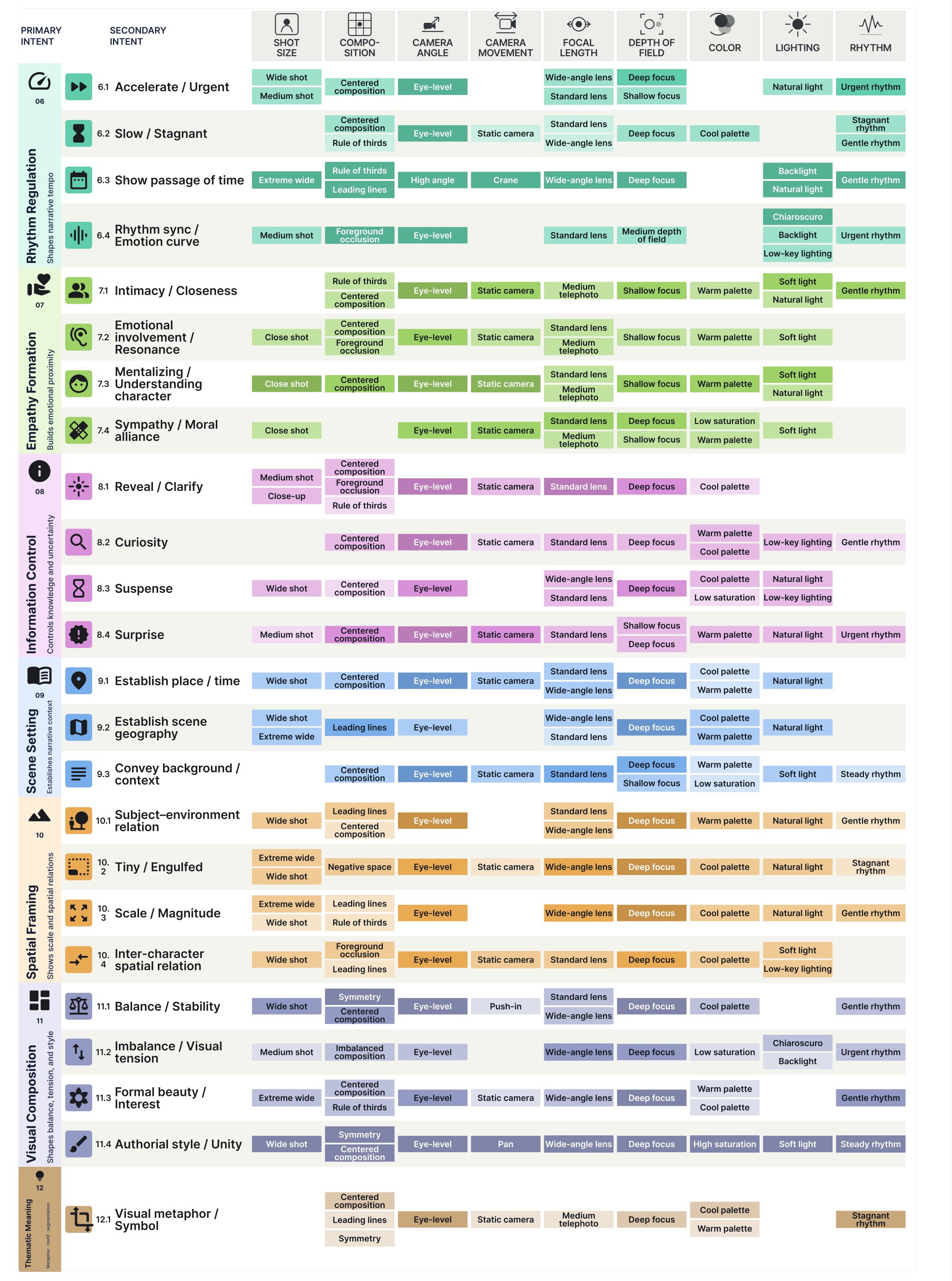}
    \caption{Design Space.
}

    \label{fig:design_space_2}
\end{figure*}

Figure~\ref{fig:design_space_1} and Figure~\ref{fig:design_space_2} present the final Cinematic Language Design Space, which connects 53 secondary intents organized into 12 primary families with their empirically grounded cinematic realizations. The intent families include \emph{Emotion Elicitation}, \emph{Gaze Guidance}, \emph{Atmosphere Construction}, \emph{Viewer Alignment}, \emph{Character Portrayal}, \emph{Rhythm Regulation}, \emph{Empathy Formation}, \emph{Information Control}, \emph{Scene Setting}, \emph{Spatial Framing}, \emph{Visual Composition}, and \emph{Thematic Expression}. Because cinematic language can serve multiple communicative purposes simultaneously, the taxonomy is multi-label, allowing a shot to be associated with multiple intents.
\begin{figure*}[t]
    \centering
    \includegraphics[width=0.95\linewidth]{Figures/design_space-1.pdf}
    \caption{The complete Intent-Oriented Cinematic Language Design Space. It organizes 53 secondary intents into 12 primary intent families and maps each intent to observed values across nine cinematic parameter dimensions. Colors distinguish the intent families, while cell saturation indicates how frequently each parameter value was observed for the corresponding intent.}
    \label{fig:design_space_1}

\end{figure*}
For each secondary intent, the Design Space records the observed frequency of values across the nine cinematic parameter dimensions. Figure~\ref{fig:design_space_1} visualizes these associations, with saturation indicating the frequency of each parameter value. The resulting mappings preserve the many-to-many relationship between intents and cinematic techniques, capturing alternative ways of expressing the same intent rather than prescribing a fixed configuration. For example, \emph{Mysterious/Solitary/Detached} may be commonly realized through cool colors, a wide-angle lens, deep focus, low-key lighting, and restrained pacing.

To make this knowledge accessible throughout the system, we organize the intent definitions, representative examples, related or confusable intents, parameter values, and their associations into a shared vector index ($\mathfrak{D}$). Both intents and parameter values are embedded as searchable representations while retaining their semantic relationships. The Intent Perceiver, Strategy Recommender, Cinematic Editor, and Preference Learner all query ($\mathfrak{D}$) through retrieval-augmented generation (RAG), using retrieved knowledge as contextual evidence rather than deterministic rules. Specifically, the Intent Perceiver retrieves relevant intents for intent alignment; the Strategy Recommender retrieves intent--cinematic parameter mappings for strategy construction; the Cinematic Editor retrieves the expressive implications of relevant parameters to guide intent-oriented editing; and the Preference Learner retrieves relevant cinematic concepts to interpret recurring user choices. Thus, the Design Space functions not only as a representation of cinematic language but also as a shared, retrievable knowledge substrate that connects creators' intents with actionable cinematic decisions throughout the creation process.

%% file: Sections/06-Agents.tex

\section{The Multi-Agent System}
\label{sec:multi_agent_engine}

Building on the Cinematic Language Design Space, \name integrates three complementary agents: (1) \emph{Intent Perceiver}, (2) \emph{Strategy Recommender}, and (3) \emph{Preference Learner} to support intent understanding, cinematic strategy development, and adaptive co-creation. Together, they form the core reasoning engine of \name. We detail these three agents in the following sections.

\subsection{Intent Perceiver}
\label{sec:intent-perceiver}
The \emph{Intent Perceiver} translates underspecified creative requests into a user-confirmed \emph{Intent Brief} grounded in the Cinematic Language Design Space. Given a reference image $r_t$ and a natural-language request $q_t$, it interprets the intended effect in context and retrieves relevant secondary intents, definitions, related concepts, and associated cinematic parameters from the indexed Design Space $\mathfrak{D}$. The retrieved knowledge grounds the generation and evaluation of candidate intent hypotheses. When multiple interpretations could lead to substantially different cinematic realizations, the agent initiates targeted dialogue in the \emph{Dialog Box}, presenting questions or alternative interpretations to resolve the ambiguity; otherwise, it proceeds directly to confirmation. The agent then synthesizes the selected intents, contextual interpretation, desired expressive effect, and remaining assumptions into an \emph{Intent Brief} $B_t$, which the creator can inspect and correct. Only after explicit confirmation is $B_t$ passed to the \emph{Strategy Recommender}, ensuring that subsequent cinematic strategies are grounded in a creator-approved interpretation.

\subsection{Strategy Recommender}
\label{sec:strategy_recommender}

The \emph{Strategy Recommender} translates the confirmed Intent Brief into a small set of alternative cinematic realizations. It consists of a \emph{Strategy Generator}, which retrieves relevant cinematic parameter values and composes them into candidate strategies, and a \emph{Strategy Reviewer}, which validates and repairs the generated candidates. The process comprises three stages: \emph{parameter selection}, \emph{strategy generation}, and \emph{strategy review and repair}.

\paragraph{\textbf{Parameter Selection.}}
Given the confirmed Intent Brief $B_t$, the Strategy Generator identifies its confirmed secondary intents and retrieves their associated cinematic parameter values from the shared Design Space index $\mathfrak{D}$. Let $\mathcal{Y}$ denote the 53 secondary intents in the Cinematic Language Design Space. The confirmed intents and their retrieved parameter values are:
\begin{equation}
\label{eq:parameter-pool}
Y_t
=
\{y\in\mathcal{Y}\mid y\text{ is confirmed in }B_t\},
\qquad
\mathcal{A}_t
=
\operatorname{Retrieve}(\mathfrak{D},Y_t),
\end{equation}
where $\mathcal{A}_t$ contains the parameter values associated with the confirmed intents. These retrieved associations serve as candidate choices rather than fixed prescriptions. The Generator further filters them according to the reference image $r_t$ and the contextual constraints in $B_t$, excluding values that are infeasible or poorly suited to the current visual context. The resulting $\mathcal{A}_t$ provides the candidate parameter space for strategy generation.

\paragraph{\textbf{Strategy Generation.}}
The Generator uses LLM-based reasoning to compose compatible parameter values from $\mathcal{A}_t$ into distinct cinematic realizations of the confirmed intents. Rather than applying the retrieved parameters directly, it reasons over their relationships to construct alternative combinations that express the same intents through different cinematic treatments. When multiple intents are confirmed, each strategy jointly considers their requirements. The Generator then organizes each realization into a temporally coherent shot sequence, producing:
\begin{equation}
\label{eq:strategy-generation}
S_t^{(0)}
=
\left\{
s_i^{(0)}
=
\left(
\sigma_i,
q_{i1},\ldots,q_{in_i}
\right)
\right\}_{i=1}^{K},
\qquad K\leq3,
\end{equation}
where $\sigma_i$ describes the overall cinematic strategy and $q_{ij}$ denotes the parameter configuration of its $j$-th shot. Each $q_{ij}$ specifies the relevant cinematic parameters, shot duration, transition to the preceding shot, and instructions for adapting the design to the reference image.

\paragraph{\textbf{Strategy Review and Repair.}}
The Strategy Reviewer evaluates each candidate for \emph{validity}, \emph{intent alignment}, \emph{temporal coherence}, and \emph{diversity}. Rule-based checks verify parameter completeness and validity, compatibility among parameters within each shot, continuity across adjacent shots, and compliance with the overall duration constraint. LLM-based reasoning assesses whether the strategy faithfully realizes $B_t$, remains appropriate for the reference image, forms a coherent shot sequence, and provides a meaningfully distinct alternative. For candidates that fail these checks, the Reviewer returns structured feedback on the affected shots or parameters, which the Generator uses to revise the candidate. Review and repair repeat until the candidate passes the checks or reaches the predefined iteration limit. The resulting validated set $\widehat{S}_t$ is then presented in the Strategy Explorer for comparison and selection.

\begin{figure*}[t]
    \centering
\includegraphics[width=1\linewidth]{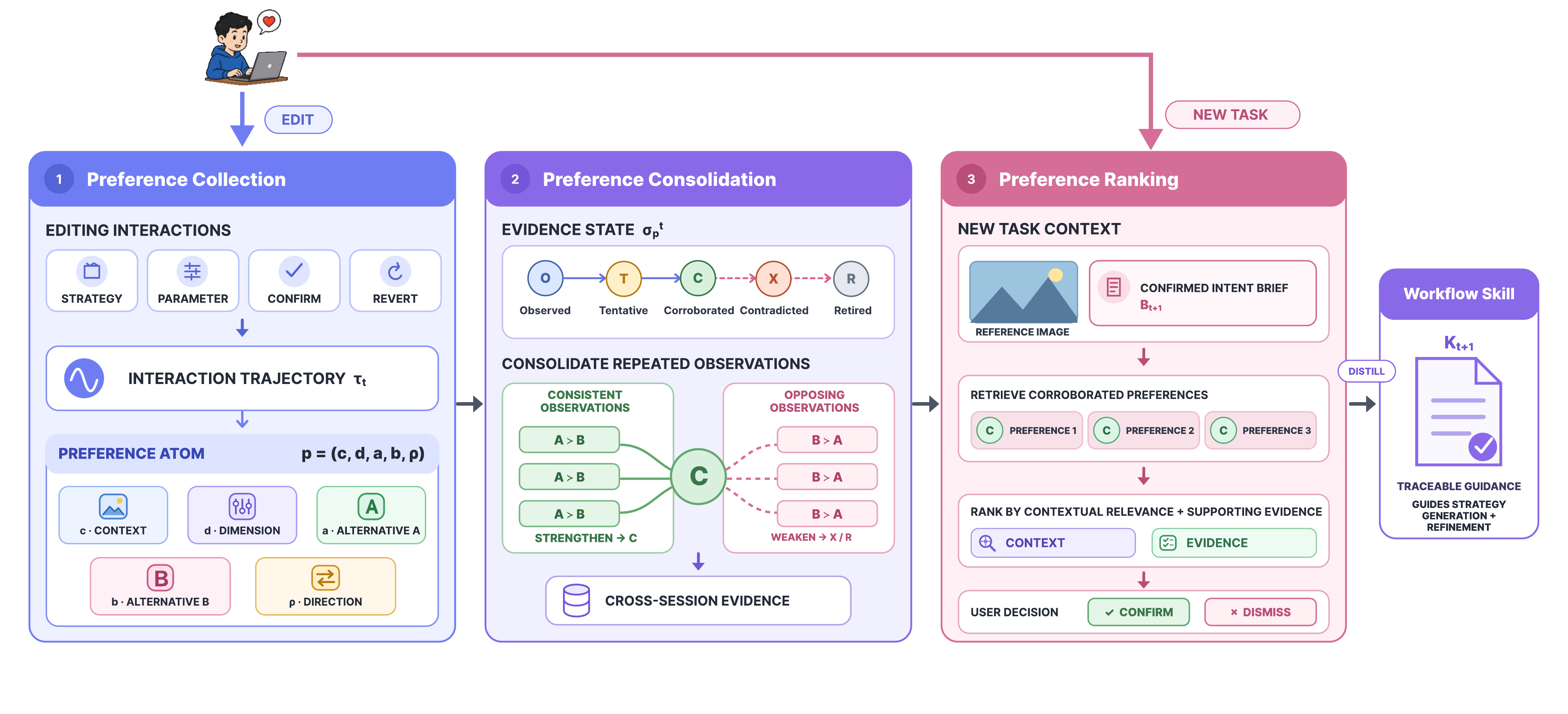}
    \vspace{-3em}
    \caption{Overview of the Preference Learner, which operates in three stages. (1) Preference Collection converts editing interactions into contextualized preference atoms. (2) Preference Consolidation integrates consistent and opposing evidence across sessions to update preference states. (3) Preference Ranking retrieves and ranks applicable preferences for a new task; user-confirmed preferences are then distilled into a workflow skill that guides strategy generation and refinement.}
    \label{fig:preference}
    \Description{Three-stage Preference Learner workflow. Preference Collection converts strategy selections, parameter edits, confirmations, and reversions into contextualized preference atoms containing context, cinematic dimension, two alternatives, and preference direction. Preference Consolidation combines consistent and opposing observations across sessions and updates each preference through observed, tentative, corroborated, contradicted, and retired states. For a new task, Preference Ranking retrieves corroborated preferences, ranks them by contextual relevance and supporting evidence, and asks the user to confirm or dismiss them. Confirmed preferences are distilled into a workflow skill that guides strategy generation and refinement.}
    \vspace{-2.5em}
\end{figure*}

\subsection{Preference Learner}
\label{sec:preference-learner}

The \emph{Preference Learner} learns reusable, context-sensitive cinematic preferences from users' interaction histories. Rather than treating an individual choice as a stable preference, it accumulates evidence from repeated decisions made for similar intents and visual contexts. For a new task, relevant preferences are retrieved and presented to the user for confirmation before they influence subsequent strategy generation. As shown in Figure~\ref{fig:preference}, the Preference Learner consists of three stages: \emph{Preference Collection}, \emph{Preference Consolidation}, and \emph{Preference Ranking}.

\paragraph{\textbf{Preference Collection.}}
Given an interaction trajectory $\tau_t$, the Learner extracts preference evidence from strategy selections, parameter edits, confirmations, and reversions. To make directional observations
comparable across sessions, each is represented as a
\emph{preference observation}:
\begin{equation}
\label{eq:preference-atom}
p=(c,d,a,b,\rho),
\end{equation}
where $c$ denotes the context defined by the user's intended effect and the visual content of the reference image, $d$ denotes the cinematic parameter dimension involved in the decision (e.g., composition or camera movement), $a$ and $b$ denote the compared parameter values or strategies, and
$\rho\in\{a\succ b,\;b\succ a\}$ records the observed preference direction. For example, when the user intends to convey \emph{loneliness} in an image showing a character alone in a large environment, choosing \emph{negative space} over \emph{centered composition} along the \emph{composition} dimension records a preference
$\rho=\textit{negative space}\succ\textit{centered composition}$ under that context. Such a preference is therefore associated with the specific intent and visual context rather than interpreted as a general preference for negative space.

\paragraph{\textbf{Preference Consolidation.}}
The Learner accumulates preference observations across sessions by matching those that share a similar intended effect, visual context, cinematic parameter dimension, and alternatives. A preference is not considered stable after a single observation; instead, its evidence state $\sigma_p^t$ evolves as new observations accumulate:
\begin{equation}
\label{eq:preference-states}
\sigma_p^t
\in
\{\mathsf{O},\mathsf{T},\mathsf{C},\mathsf{X},\mathsf{R}\},
\end{equation}
where $\mathsf{O}$ (\emph{Observed}) denotes a newly observed preference with insufficient evidence, $\mathsf{T}$ (\emph{Tentative}) denotes a preference that has recurred but is not yet sufficiently supported, $\mathsf{C}$ (\emph{Corroborated}) denotes a preference supported by consistent observations across comparable contexts, $\mathsf{X}$ (\emph{Contradicted}) denotes a preference with both supporting and opposing observations, and $\mathsf{R}$ (\emph{Retired}) denotes a preference whose contradictory evidence persists and is therefore excluded from future recommendations. Consistent observations progressively move a preference from $\mathsf{O}$ to $\mathsf{T}$ and then $\mathsf{C}$, whereas repeated opposing observations move it toward $\mathsf{X}$ and eventually $\mathsf{R}$. This evidence-based progression prevents isolated or inconsistent decisions from being generalized as stable preferences.

\paragraph{\textbf{Preference Ranking.}}
For a new editing task, the Learner retrieves corroborated preferences whose context is compatible with the confirmed Intent Brief $B_{t+1}$ and reference image $r_{t+1}$:
\begin{equation}
\label{eq:applicable-preferences}
\mathcal{P}_{t+1}
=
\left\{
p \mid
\sigma_p^t=\mathsf{C},
\;
c_p\sim(B_{t+1},r_{t+1})
\right\},
\end{equation}
where $c_p\sim(B_{t+1},r_{t+1})$ indicates that the intent and visual context associated with preference $p$ are relevant to the current task. The retrieved preferences are ranked by contextual relevance and supporting evidence, and the highest-ranked ones are presented as suggestions in the \emph{Dialog Box}. The user can \emph{Confirm} a suggestion to apply the preference to the current task or \emph{Dismiss} it to reject its use. Let $\pi_p^{t+1}\in\{0,1\}$ indicate this task-level decision, where $\pi_p^{t+1}=1$ denotes user confirmation. The active preferences are therefore
\begin{equation}
\label{eq:active-preferences}
\mathcal{P}^{t+1}_{u,\mathrm{active}}
=
\left\{
p\in\mathcal{P}_{t+1}
\mid
\pi_p^{t+1}=1
\right\}.
\end{equation}

Finally, the confirmed preferences are distilled into a contextual \emph{workflow skill} that guides downstream strategy generation and refinement:
\begin{equation}
\label{eq:workflow-skill}
K_{t+1}
=
\operatorname{Distill}_{\mathcal{D}}
\left(
\mathcal{P}^{t+1}_{u,\mathrm{active}},
B_{t+1}
\right),
\end{equation}
where $\mathcal{D}$ denotes the Cinematic Language Design Space. The workflow skill translates the selected preferences into actionable guidance, such as prioritizing particular cinematic parameter values or considering specific alternatives during strategy generation and refinement. Each recommendation remains linked to its supporting preference, making the learned guidance traceable and allowing the user to accept or dismiss it without overriding the confirmed intent or subsequent explicit edits.

%% file: Sections/07-Evaluation.tex
\section{Evaluation}
\label{sec:eval}
We evaluated \name{} through two complementary studies: 
(1) a user study assessing the generated videos, and (2) a case study with expert users followed by interviews assessing the system's end-to-end usability. All studies involving human participants received IRB approval, and all participants provided written informed consent prior to participation.

\subsection{Evaluation of Generated Videos}

\begin{figure*}[t]
    \centering
    \includegraphics[width=1\linewidth]{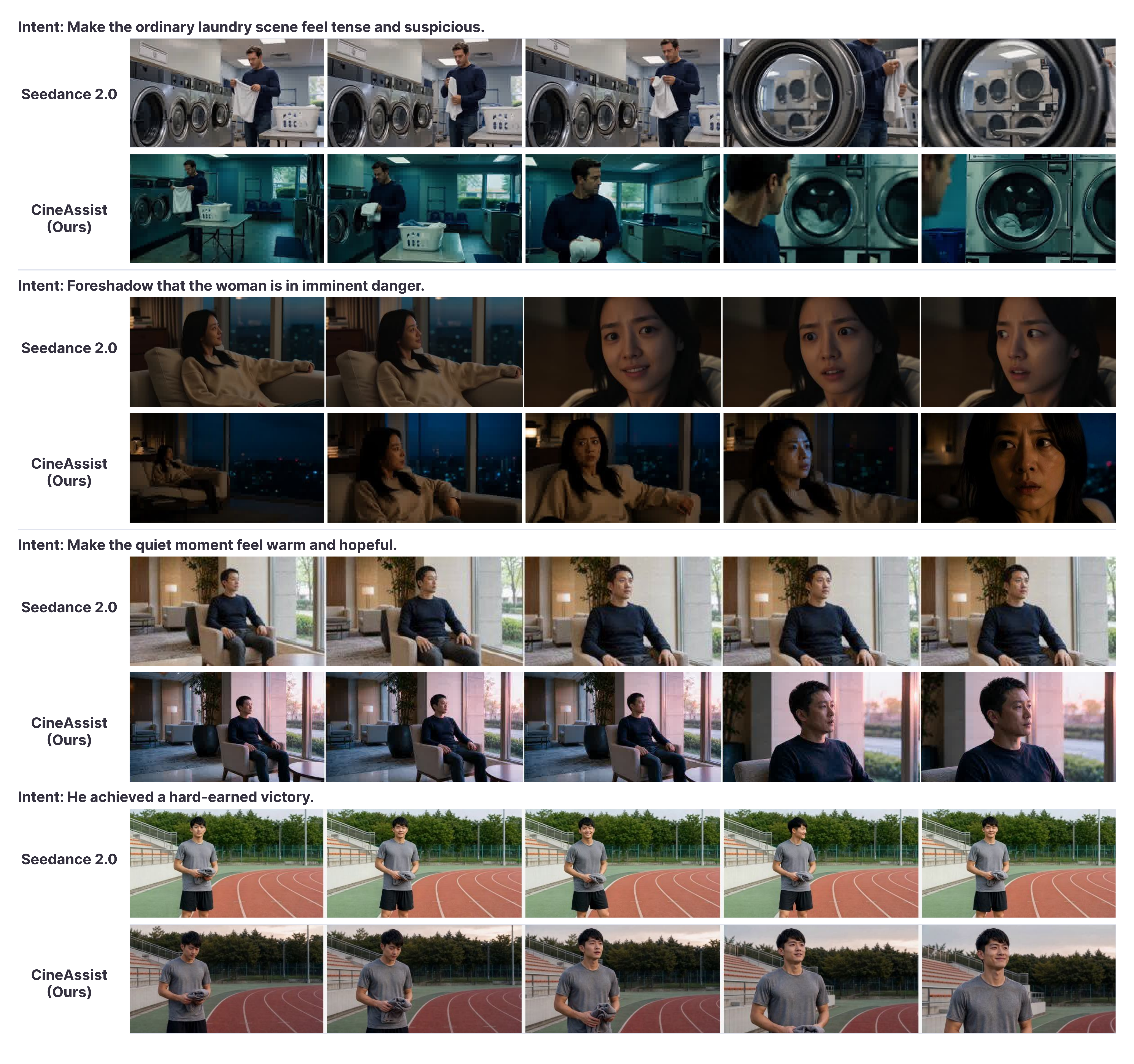}
    \caption{Qualitative comparison between \name and direct Seedance~2.0 generation on four representative test cases. Five temporally ordered keyframes are shown for each condition and intent.}
    \Description{Four qualitative comparisons between direct Seedance 2.0 generation and CineAssist. Each case contains two rows of five temporally ordered keyframes for the same reference image and intent. The intents concern making a laundry scene tense and suspicious, foreshadowing danger to a woman, making a quiet moment warm and hopeful, and depicting a hard-earned victory. The paired sequences visually compare how the two methods vary framing, camera distance, viewpoint, movement, lighting, and progression over time.}
    \label{fig:cases}
\end{figure*}
We conducted a questionnaire-based user study to evaluate whether the videos generated through \name{} communicated the target intents more effectively than those generated directly by the underlying video generation model.

\paragraph{\textbf{Stimuli.}}
We compared \name{} against Seedance~2.0~\citep{seedance2026seedance20}, an accessible state-of-the-art video generation model. To ensure a fair comparison, both conditions used the same reference image, underspecified target intent, and underlying video generation model. In the \name{} condition, the reference image and intent were first processed through our intent-alignment and cinematic-strategy pipeline to produce a Shot Script, which was then used to generate the video with Seedance~2.0. In the baseline condition, the reference image and intent were provided directly to Seedance~2.0 through its Omni Reference mode, without intermediate intent alignment or cinematic-strategy generation.

We prepared 15 test cases, each comprising a reference image and an underspecified target intent selected to cover diverse communicative effects represented in the cinematic language design space. For each case, we generated one video under each condition, resulting in 30 videos in total. Videos within each pair were matched in duration, ranging from 5 to 15 seconds.

\paragraph{\textbf{Participants and procedure.}}
We recruited 30 participants (17 female, 13 male; mean age = 25.7). Each participant evaluated all 15 video pairs, with each pair consisting of one video generated by \name{} and one generated by the baseline. To minimize presentation bias, the two videos were labeled only as Video A and Video B, and the assignment of methods to A or B was counterbalanced across pairs. After viewing each pair, participants rated the two videos on three 7-point comparison scales: (1) clarity of communication of the target intent, (2) appropriateness of the cinematic language, and (3) overall preference for the final result. For each scale, 1 indicated a strong preference for Video A, 7 indicated a strong preference for Video B, and 4 indicated no preference.

\begin{figure}[width=\columnwidth,pos=ht]
  \centering
  \includegraphics[width=1\columnwidth]{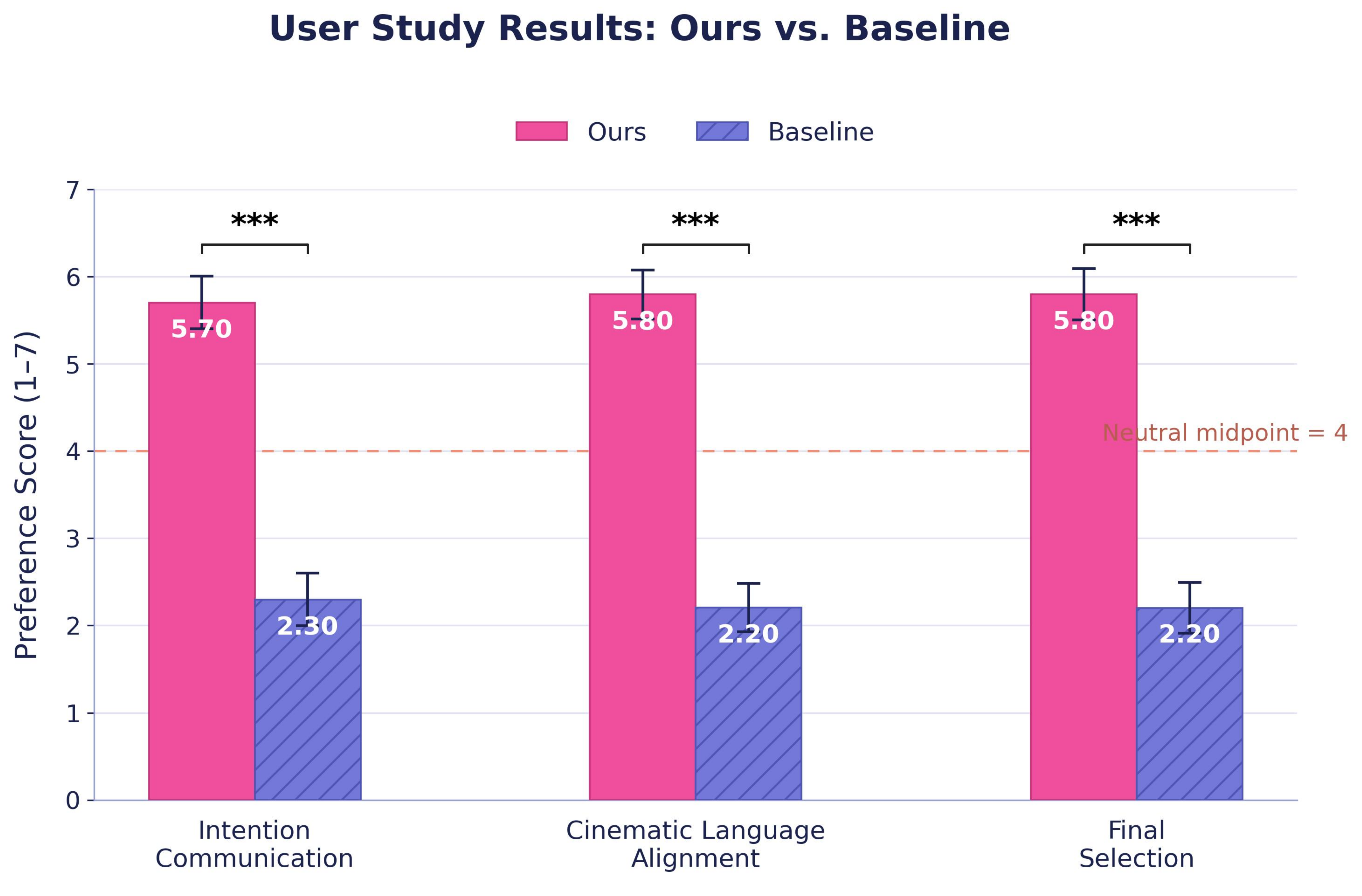}
  \caption{Comparison of \name{} and \textit{Seedance2.0}. Error bars show 95\% confidence intervals ($^{***}p_{\mathrm{Holm}}<.001$).}
  \Description{Grouped bar chart comparing CineAssist with direct Seedance 2.0 generation on a seven-point preference scale, with four as the neutral midpoint. CineAssist receives mean scores of 5.70 for intent communication, 5.80 for cinematic-language alignment, and 5.80 for final selection. The corresponding baseline scores are 2.30, 2.20, and 2.20. Error bars show 95 percent confidence intervals, and all three differences are statistically significant after Holm correction.}
  \label{fig:generated-results-comparison}
\end{figure}

\paragraph{\textbf{Analysis and Results.}}
For each question, we averaged each participant's ratings across the 15 video pairs and tested the resulting participant-level means against the neutral value of 4 using one-sided Wilcoxon signed-rank tests~\citep{wilcoxon1945}. Holm correction~\citep{holm1979} was applied across the three questions.

Participants significantly preferred \name{} across all three measures. Compared with the baseline, \name{}'s videos were judged to communicate the target intent more clearly, use cinematic language more appropriately, and be more suitable as final results (some representative cases are shown in Figure~\ref{fig:cases}). All three comparisons remained significant after Holm correction ($p_{\mathrm{Holm}}<.001$), with large effect sizes. The preference was also consistent at the individual-pair level, with more than four-fifths of responses favoring \name{} on each measure (Figure~\ref{fig:generated-results-comparison}). Together, these results suggest that \name{} can more effectively translate underspecified creative intents into cinematic designs that are perceived as clearer, more appropriate, and more preferable.

\subsection{Evaluation of System Usability}
\label{sec:eval-evolution-user}

We conducted an expert study to examine \name{} from two complementary perspectives: its end-to-end support for cinematic creation and its ability to adapt to users' preferences. The study first asked experts to use the complete \name{} workflow to create videos from their own creative intents. This case study allowed us to examine how experts experienced intent alignment, strategy exploration, and interactive cinematic refinement in practice. The interactions collected during these tasks were subsequently used to evaluate whether the Preference Learner could transfer users' preferences to new creation contexts. Semi-structured interviews further elicited participants' perceptions of the system's usefulness, ease of use, and limitations.

\paragraph{\textbf{Participants.}}
We recruited six experts with complementary backgrounds in cinematic production and generative AI. Three film professionals (P1--P3; 5--8 years of experience; 2 female; mean age: 28) worked as directors or writer-directors and had extensive expertise in cinematic language and shot design. Three AIGC practitioners (A1--A3; 3--5 years of experience; 1 female; mean age: 26) primarily used generative AI to create visual assets and had substantial experience with AI-assisted content creation, but less formal training in cinematic language and shot design. This grouping allowed us to examine \name{} from both cinematic-production and generative-AI perspectives.

\paragraph{\textbf{Procedure.}}
The study was conducted remotely via an online meeting platform. Each session began with a 10-minute introduction to \name{}'s core functions and workflow, followed by independent exploration. Participants then completed five end-to-end creation tasks using reference images and their own creative intents. For each task, they described their intended effect, refined the intent through dialogue, confirmed an image-specific Intent Brief, explored alternative cinematic strategies, selected and refined a strategy in the Cinematic Editor, and generated a final video. This process allowed participants to experience the complete workflow from intent specification to video generation.

After each task, participants rated the Intent Brief, strategy alternatives, and editing controls using five-point Likert scales. Their selections, edits, and other preference-relevant interactions were recorded as evidence for subsequent preference learning. After completing the five creation tasks, participants rated the overall usefulness and ease of use of \name{} and completed a semi-structured interview about their experience, including perceived benefits, challenges, and limitations.

To further examine preference adaptation, we conducted a follow-up evaluation using six new reference images. Participants did not operate the system in this phase. For each image, they compared two videos generated under identical settings, with or without the preferences learned from their creation interactions. The videos were presented in randomized order without condition labels. Participants selected their preferred video and rated its preference fit and fidelity to the intended effect.

\begin{figure*}[t]
    \centering
    \includegraphics[width=0.98\linewidth]{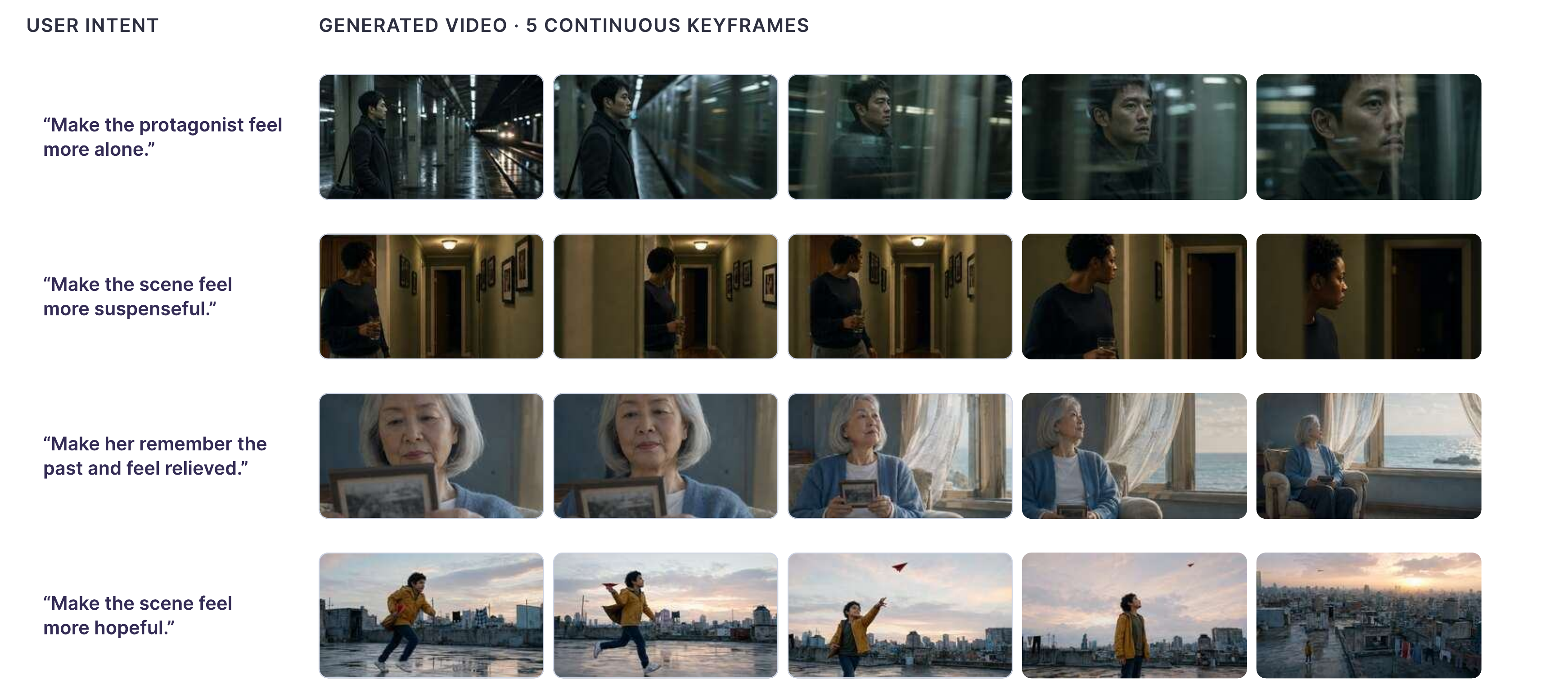}
    \caption{Four representative videos created with the complete CineAssist workflow during the end-to-end creation phase.}
    \label{fig:case}
    \Description{Four videos produced by expert users with CineAssist, each represented by five consecutive keyframes. The first sequence uses reflections and progressively closer views to make a man feel more alone. The second follows a man approaching a dark doorway to increase suspense. The third gradually widens from an older woman holding a photograph to reveal a bright seaside setting, expressing remembrance and relief. The fourth follows a person releasing a paper airplane and ends with a wide sunset city view to communicate hope}
\end{figure*}
\begin{figure}[width=\columnwidth,pos=ht]
    \centering
    \includegraphics[width=1\linewidth]{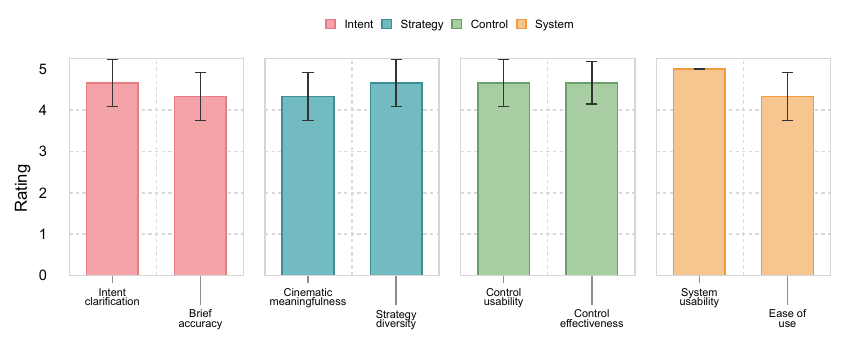}
    \caption{Ratings of intent alignment, strategy exploration, editing control, and the overall system across different criteria on a 5-point Likert scale, as provided by 6 expert users, where 5 is the best and 1 is the worst. Error bars represent standard deviations.}
    \Description{Bar charts of six expert users’ ratings on eight criteria using a five-point scale. The criteria cover intent clarification, Intent Brief accuracy, cinematic meaningfulness, strategy diversity, control usability, control effectiveness, system usability, and ease of use. Mean ratings are all above four.}
    \label{fig:end-to-end-ratings}
\end{figure}
\paragraph{\textbf{Results.}}
Figure~\ref{fig:case} presents representative videos created by the expert users during the end-to-end creation tasks, while Figure~\ref{fig:end-to-end-ratings} summarizes their questionnaire ratings. Overall, \name{} received positive ratings across the evaluated dimensions. The interviews further revealed three recurring themes concerning the end-to-end creation experience, followed by evidence of preference adaptation in the follow-up evaluation.

\underline{\textit{Effective Intent Alignment.}}
Participants found that the intent-alignment process progressively concretized underspecified intents while preserving their control over the interpretation~(P1, P3, A1--A3). Structured questions and selectable interpretations helped them clarify narrative emphasis, emotional tone, subject relationships, and desired audience response. P3 explained, \emph{``It did not decide the meaning for me. It offered several possible interpretations, and I could choose the one closest to my intent.''}

\underline{\textit{Effective Parameter Mapping and Intuitive Editing.}}
Participants valued how \name{} translated high-level intents into concrete cinematic parameters~(P2--P3, A1--A3). As A2 explained, \emph{``I only needed to describe the effect I wanted, and the system turned it into specific choices of cinematic parameters.''}
Participants across both groups also spoke highly of the 3D Cinematic Editor. Film professionals noted that directly positioning the camera in the 3D workspace resembled shot arrangement in filmmaking practice, making the editing process familiar and convenient~(P1--P3). As P1 explained, \emph{`It feels similar to arranging a shot on set. I can directly position the camera and see how the framing changes.''} AIGC practitioners found this approach more intuitive than repeatedly revising prompts~(A2--A3). A2 commented, \emph{`Directly moving the camera is much clearer than repeatedly changing the prompt and waiting to see whether it works.''} Participants with less formal cinematic training also valued the intent-driven editing guidance, which helped them understand whether parameter changes supported their intended effects~(A1--A3). As A1 explained, \emph{``The timely guidance helps me adjust the cinematic parameters because I am not always sure whether my changes are appropriate for the effect I want.''}

\underline{\textit{Integrated and Professional Workflow.}}
Participants appreciated how \name{} integrated intent clarification, cinematic strategy exploration, direct editing, and video generation into a coherent workflow. AIGC practitioners found the system accessible and valued its ability to support cinematic creation despite limited formal training in cinematic language. As A3 noted, \emph{``Even though I am not very familiar with cinematic language, the system helped me successfully create a clip that clearly expressed my intent.''} Film professionals, meanwhile, considered the workflow consistent with professional cinematic practice. P2 explained, \emph{``Its approach to shot design and editing reflects professional filmmaking practice rather than simply generating a visually appealing result.''} Overall, participants considered \name{} useful and relatively easy to use~(P1--P3, A1--A3).

\underline{\textit{Adaptive Preference Learning.}}
The interactions collected during the creation tasks also provided evidence for evaluating preference adaptation. Across six new reference images, the Preference Learner achieved 77.8\% preference-match accuracy when transferring learned preferences to new visual contexts. Participants also valued the ability to reuse recurring cinematic preferences without repeatedly specifying the same choices. A1 explained, \emph{``When I create a longer sequence, I usually want the shots to maintain a consistent style. Remembering my preferences reduces the work of specifying the same choices repeatedly.''} P1 further emphasized the importance of retaining control over preference application: \emph{``A consistent visual language is important across shots, but each scene may serve a different narrative purpose. The system lets me decide whether to apply a learned preference.''}

\subsection{Example Case}
We present a representative case from the expert user study (Section~\ref{sec:eval-evolution-user}) to illustrate how \name{} helps creators transform a simple intent into a cinematically expressive video.

\begin{figure*}[t]
    \centering
    \includegraphics[width=1.0\linewidth]{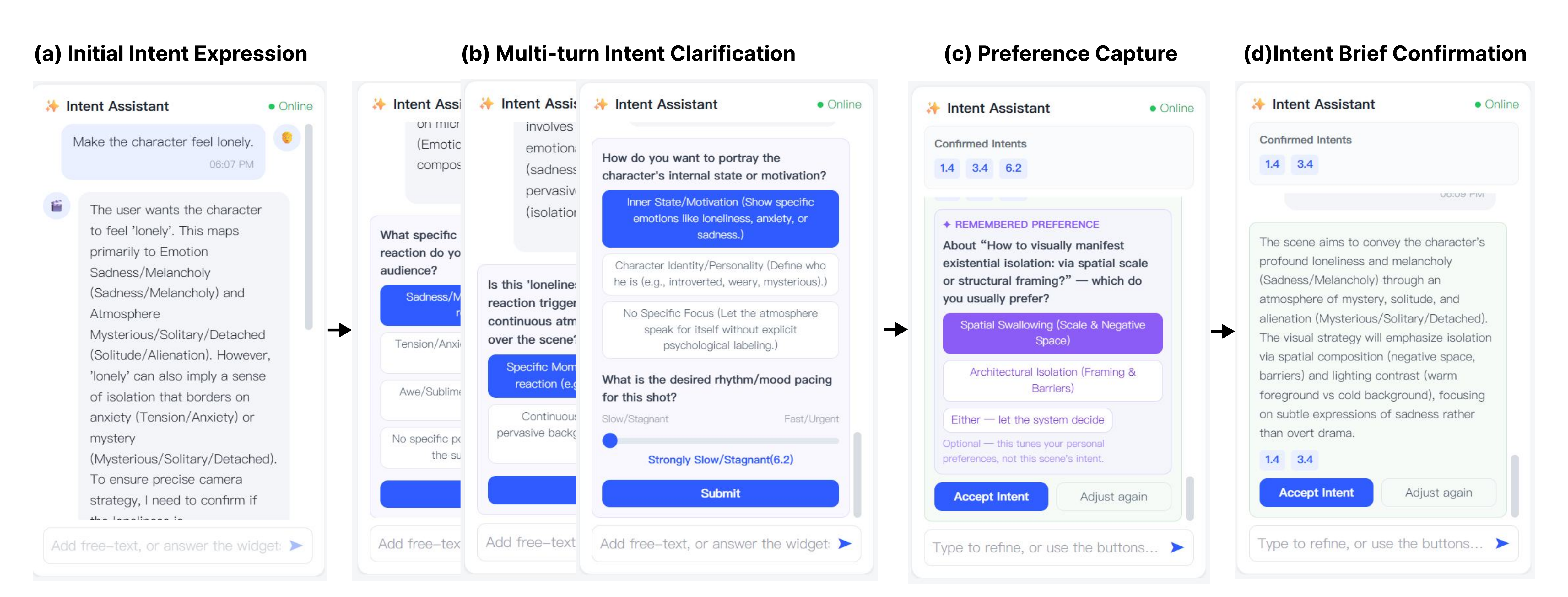}
    \caption{Clarifying an underspecified intent before cinematic strategy generation. Starting from the creator's initial request, \name{} progressively clarifies the intended experience through structured multi-turn interaction, captures the creator's preferred visual expression, and synthesizes the resulting decisions into a confirmed Intent Brief.}
    \label{fig:case_intent}
    \Description{Interface screenshots showing progressive intent clarification. The sequence begins with the request “Make the character feel lonely,” followed by structured questions about the intended emotional experience and cinematic expression. CineAssist then presents a learned visual preference for optional confirmation and summarizes the resulting interpretation in an Intent Brief. The panels show how an underspecified request becomes a user-confirmed representation before strategy generation.}
\end{figure*}
\paragraph{Stage 1: Clarifying Intent and Exploring Cinematic Strategies.}
The creator began with the underspecified request, \emph{``Make the character feel lonely''}~(Figure\ref{fig:case_intent}~(a)). Rather than immediately translating this request into a video prompt, \name{} first interpreted the possible emotional and atmospheric meanings of loneliness and then initiated a multi-turn clarification process~(Figure\ref{fig:case_intent}~(b)). Through structured questions, the creator specified the desired audience response, the source and manifestation of the character's internal state, and the intended rhythm and pacing of the shot. Based on these responses, \name{} progressively narrowed the initial request into a more explicit account of the intended emotional experience. If the creator had completed several previous creation tasks, the Preference Learner retrieved a relevant learned preference and presented it for confirmation, allowing the creator to decide whether to apply it in the current task~(Figure~\ref{fig:case_intent}~(c)). \name{} then synthesized the clarified intent and any confirmed preference into an Intent Brief for the creator to review and confirm~(Figure~\ref{fig:case_intent}~(d)). Based on the confirmed Intent Brief, \name{} generated several cinematic strategies that translated the creator's intent into cinematic-language parameters, from which the creator selected the preferred strategy for further refinement.

\begin{figure*}[t]
    \centering
    \includegraphics[width=0.95\linewidth]{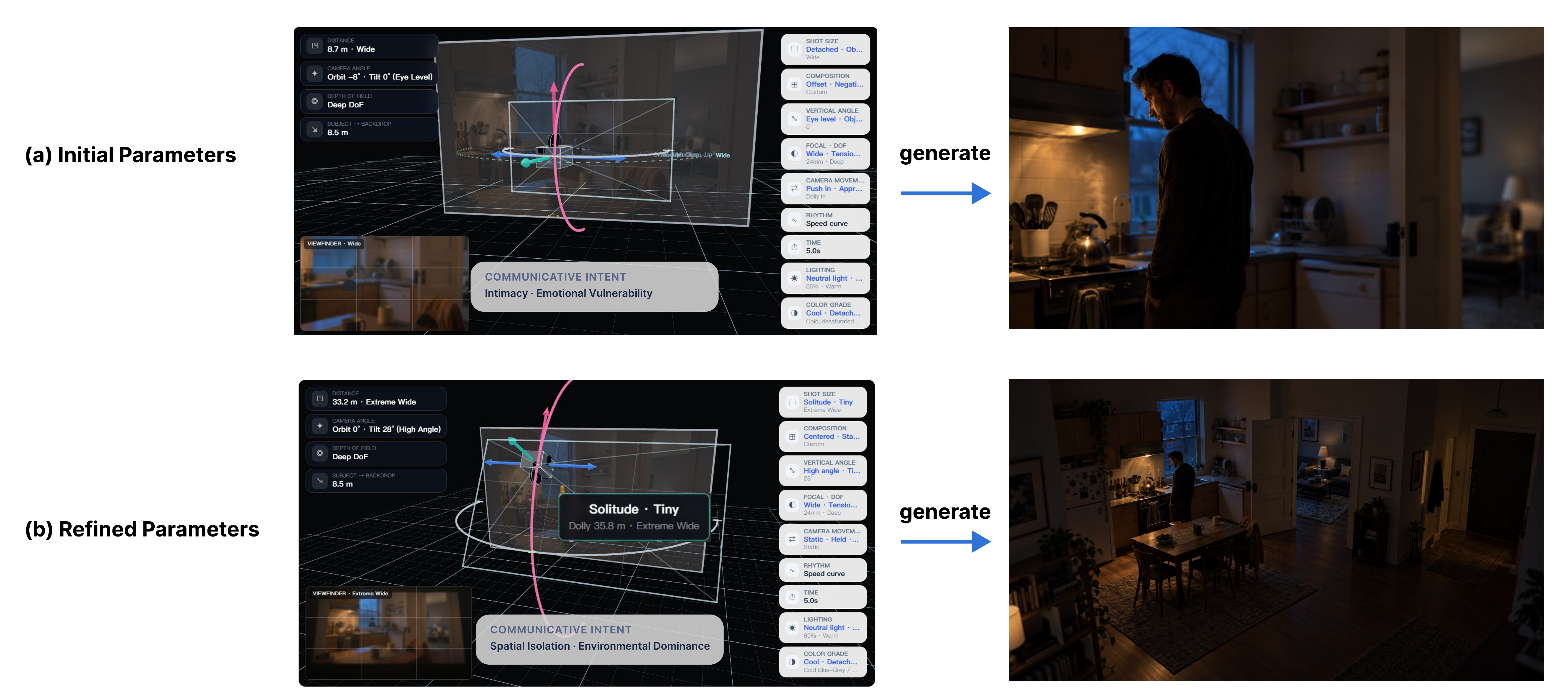}
    \caption{Refining cinematic parameters and previewing their effects on the communicative intent and generated result.}
    \Description{Before-and-after comparison of cinematic parameter refinement for expressing loneliness. The initial configuration uses a relatively close, eye-level view that emphasizes intimacy and emotional vulnerability. In the refined configuration, the camera is moved farther away and raised to form a high-angle extreme-wide view with the character occupying a small portion of the surrounding apartment. The updated intent summary emphasizes spatial isolation and environmental dominance, and the generated frames reflect this shift.}
    \label{fig:case_edit}
   
\end{figure*}

\paragraph{Stage 2: Inspecting and Refining Cinematic Parameters.}
After selecting a cinematic strategy, the creator inspected its parameter configuration in the Cinematic Editor. The initial design used a relatively close, eye-level framing that foregrounded the character, which \name{} summarized as conveying \emph{Intimacy} and \emph{Emotional Vulnerability} (Figure~\ref{fig:case_edit}~(a)). However, the creator wanted to express the character's loneliness more strongly through his relationship with the surrounding space. The creator therefore moved the camera farther away and raised its vertical angle, gradually transforming the framing into a high-angle extreme-wide shot. Throughout this process, the shot preview in the lower-left corner provided an approximate view of the resulting frame before generation, reducing unnecessary trial-and-error generation. Meanwhile, an in-situ prompt explained that moving the camera outward would make the character appear smaller and strengthen the impression of \emph{Solitude}, helping the creator understand the expressive consequence of the adjustment. Once the edit was completed, \name{} updated the communicative-intent summary from \emph{Intimacy} and \emph{Emotional Vulnerability} to \emph{Spatial Isolation} and \emph{Environmental Dominance} (Figure~\ref{fig:case_edit}~(b)). The revised framing thus shifted the visual emphasis from the character's immediate emotional state to his smallness within the surrounding apartment. The creator then clicked the generation button to generate the corresponding keyframes and final video.

\paragraph{Stage 3: Learning and Reusing Preferences across Tasks.}

\begin{figure*}[t]
    \centering
    \includegraphics[width=0.95\linewidth]{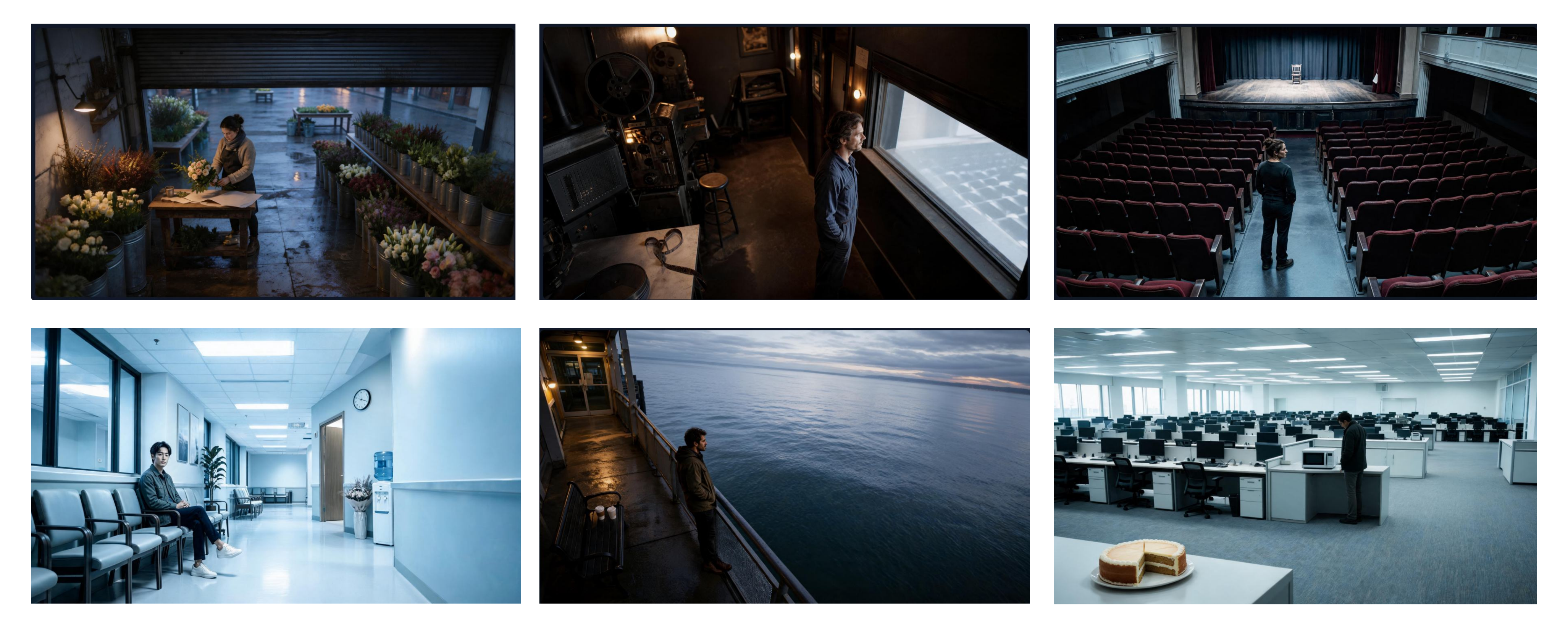}
    \caption{Preference-informed video generation across tasks. Based on cinematic preferences learned from the creator’s prior selections and refinements, \name{} adapts its strategies to diverse scenes while maintaining a consistent visual language characterized by environment-dominant framing, relatively small off-center characters, and substantial negative space.}
    \Description{Six video sequences demonstrating the reuse of a learned cinematic preference across new tasks. The scenes include a flower shop, projection room, theater, hospital, ship, and office. Although their settings differ, the generated results consistently place a relatively small, often off-center character within a large surrounding environment and preserve substantial negative space. The examples show contextual adaptation of a recurring visual preference rather than repetition of one fixed camera configuration.}
    \label{fig:preference_case}
\end{figure*}

After the creator completed several creation tasks, \name{} summarized recurring choices and refinements into reusable cinematic preferences. In this example, the creator repeatedly favored wide, environment-dominant compositions in which a solitary character occupied a relatively small portion of the frame, often positioned off-center and surrounded by substantial negative space. \name{} subsequently incorporated this preference when generating cinematic strategies for new reference images. As shown in Figure~\ref{fig:preference_case}, although the scenes varied considerably in setting, including a flower shop, projection room, theater, hospital, ship, and office, the generated results consistently preserved the creator's preferred visual language by emphasizing spatial context, isolation, and restrained composition. Rather than reproducing a fixed camera configuration, \name{} adapted the learned preference to the spatial and narrative characteristics of each new scene. The creator could inspect the applied preference and retain, revise, or override it when the intent of a particular task required a different treatment. \name{} learns the creator's preferences from prior interactions and contextually applies them to subsequent video-generation tasks.

%% file: Sections/08-Discussion.tex
\section{Discussion}

Our studies suggest that \name can help creators bridge the gap between high-level creative intents and executable cinematic designs. Beyond the specific system, our work raises broader questions about how human-agent systems can support creative tasks in which users know what they want to express but cannot easily specify how to realize it. We discuss the limitations of our current design and derive broader implications for human-agent co-creation.

\subsection{Limitations and Future Directions}
Despite the positive feedback, the expert users also raised several limitations that worth mentioning: 

\paragraph{\textbf{From Shot-Level Authoring to Long-Form Creation.}}
CineAssist currently focuses on shot-level and short-sequence authoring, which allows us to make the mapping between intent, cinematic strategy, and parameters explicit and editable. However, filmmaking involves higher-level concerns such as narrative structure, continuity, pacing across scenes, and coordination among production roles. Extending intent-driven interaction to longer sequences therefore remains an important direction. Rather than replacing existing production tools, future systems could position CineAssist as an intent-aware layer that connects shot-level cinematic reasoning with storyboarding, video generation, and professional editing workflows.

\paragraph{\textbf{From Structured Intents to Situated Creative Knowledge.}}
The Cinematic Language Design Space provides a structured vocabulary for connecting creative intents with cinematic techniques, but no fixed representation can capture the full diversity and context-dependence of cinematic expression. Different creators, genres, and cultural contexts may interpret the same cinematic technique differently. Similarly, agent-generated interpretations may occasionally misalign with a creator's intended meaning. Future work could therefore incorporate richer sources of creative evidence, including film references, sketches, visual examples, and creator-provided demonstrations, allowing the design space and agent reasoning to remain more situated and adaptable.

\subsection{Implications for Human-Agent Co-Creation}

Our work suggests a broader design pattern for human-agent creative systems: \emph{agents should mediate between human intents and executable representations rather than replace creative decisions}. In many domains, users can articulate what they want without knowing how to realize it through domain-specific operations. Making this gap an explicit interaction layer allows agents to contribute domain knowledge while keeping creative decisions under human control.

\name illustrates three principles for realizing this pattern. First, \textbf{ground creative dialogue in an explicit intermediate representation}. The Cinematic Language Design Space gives users and agents a shared vocabulary for interpreting ambiguous intents and reasoning about possible realizations. Second, \textbf{make agent reasoning inspectable and revisable}. Instead of producing a single opaque output, the system exposes alternative strategies and their underlying design parameters, enabling creators to compare, modify, and reject suggestions. Third, \textbf{let agents learn from interaction without silently taking over}. User selections and revisions provide evidence for preference adaptation, while learned preferences remain contextual and user-governed rather than becoming invisible constraints.

Together, these principles suggest a shift from \emph{generation-centered} to \emph{decision-centered} human-agent co-creation. The role of the agent is not simply to generate an artifact, nor to autonomously determine the best creative solution. Instead, it helps structure an otherwise difficult design space: clarifying what the creator means, proposing plausible ways to realize it, explaining the consequences of alternative choices, and remembering recurring preferences when appropriate. This division of labor allows human creators to retain authorship while benefiting from the agent's ability to reason over complex domain knowledge.

More broadly, we view the intermediate design representation as a potentially reusable foundation for human-agent co-creation beyond cinematography. Similar representations could be constructed in other creative domains where a semantic gap exists between \emph{what users want to express} and \emph{the domain-specific operations needed to express it}. The key lesson is not to automate this translation away, but to make the translation itself an interactive and inspectable object of collaboration.

%% file: Sections/09-Conclusion.tex
\section{Conclusion}
In this paper, we presented CineAssist, an intent-driven cinematic authoring system that helps creators translate expressive goals into cinematic designs. Through a formative study, we identified a key challenge in AI-assisted video creation: creators often understand what they want to communicate but struggle to translate those intentions into cinematic decisions.
To address this challenge, we introduced a cinematic language design space that connects creative intentions with cinematic techniques and developed an interactive workflow that supports intent alignment, strategy exploration, semantic refinement, and iterative generation. Rather than requiring users to directly manipulate low-level parameters, CineAssist enables creators to explore alternative cinematic strategies and understand how technical choices contribute to expressive outcomes.
Our work demonstrates a shift from parameter-centered control toward intent-driven human-AI co-creation, where AI serves not as an autonomous decision maker but as a creative partner that helps users articulate, explore, and refine cinematic expressions. We hope this perspective inspires future interactive systems that bridge the gap between human creative intent and computational media generation.